\documentclass[12pt,superscriptaddress,notitlepage,nofootinbib]{revtex4-1}
\usepackage[utf8x]{inputenc}
\usepackage[T1]{fontenc}
\usepackage{graphicx}
\usepackage{color}

\usepackage{mathtools}
\usepackage{amssymb}
\usepackage{amsthm}

\begin{document}
\title{Symmetry Breaking and Phase Transitions
in Random Non-Commutative Geometries and Related
Random-Matrix Ensembles}
\author{Mauro~D'Arcangelo}
\email{darcangelo.mauro@gmail.com,}
\affiliation{School of Mathematical Sciences, University of Nottingham, University Park, Nottingham NG7 2RD, UK}
\author{Sven~Gnutzmann}
\email{sven.gnutzmann@nottingham.ac.uk}
\affiliation{School of Mathematical Sciences, University of Nottingham, University Park, Nottingham NG7 2RD, UK}

\begin{abstract}
    Ensembles of random fuzzy non-commutative geometries
    may be described in terms of finite (\(N^2\)-dimensional)
    Dirac operators and a probability measure.
    Dirac operators of type \((p,q)\) are defined in 
    terms of  commutators and anti-commutators of \(2^{p+q-1}\)
    hermitian matrices \(H_k\)
    and tensor products with a representation of a Clifford algebra.
    Ensembles based on this idea have recently been 
    used as a toy model for 
    quantum gravity,
    and they are interesting random-matrix 
    ensembles in their own right.
    We provide 
    a complete theoretical picture of crossovers,
    phase transitions, and symmetry breaking
    in the \(N \to \infty \) limit of 
    1-parameter families of quartic Barrett-Glaser ensembles 
    in the one-matrix cases \((1,0)\) and \((0,1)\) that 
    depend on one coupling constant \(g\).
    Our theoretical results are in full agreement with previous 
    and new Monte-Carlo simulations.
\end{abstract}

\maketitle

\section{Introduction}

We consider two one-parameter families of random matrix ensembles that have
recently been introduced by Barrett and Glaser \cite{barrett2016}
in the context of random fuzzy spaces and non-commutative geometries \cite{barrett2015}.
A central object in this approach is a Dirac-operator \(D\) in a finite
dimensional Hilbert-space. The latter stores geometric information about the
underlying space just as a Laplace operator stores information about the
underlying Riemannian manifold. We will focus on the two simplest settings.
These are referred to as \((1,0)\) and \((0,1)\) geometries. In these cases the
Hilbert space has dimension \(N^2\) for some positive integer \(N\)
and 
may thus be identified as \( \mathbb{C}^{N^2}\) which we represent
%the tensor product \(\mathbb{C}^N
%\otimes \mathbb{C}^N\) 
%or
as the space of \(N \times N\)
square complex matrices
\(\mathcal{M}_{N}(\mathbb{C})\). 
A Dirac operator is then defined in terms
of a given hermitian \( N\times N\) matrix \(H\) by
\begin{subequations}
  \begin{align}
    \qquad \qquad D_{(1,0)}\equiv D_+
    &
      = H\otimes \mathbb{I} + \mathbb{I} \otimes H^T
    &
      \equiv
    &
      \left\{ H, \cdot\right\}\qquad\qquad\\
    \intertext{in the \((1,0)\) case, and }
    D_{(0,1)} \equiv D_-
    &
      = H\otimes \mathbb{I} - \mathbb{I} \otimes H^T
    &
      \equiv
    &
      \left[ H, \cdot\right]
  \end{align}
  \label{Dirac}
\end{subequations}
in the \((0,1)\) case. We will often refer 
to the \((1,0)\) case just as \(+\) and the
\((0,1)\) case by \(-\).
In the latter case the operator \(D_-\) does 
not change when \(H\) is shifted by a 
constant \(H \to H + E  \). We will thus 
assume that \(H\) has a vanishing trace 
\(\mathrm{tr}\ H=0\) in the \((0,1)\) case.
There are further types of 
geometries referred to as 
\( (p,q) \) which are defined in terms
commutators and anti-commutators of 
\(2^{p+q-1}\) hermitian matrices in a 
larger Hilbert
space that carries a representation of a Clifford algebra.
Many of these geometries are interesting as toy models both 
from a
quantum gravity and a random-matrix point of view. 
The analytical methods we apply in this work are suitable if \(D\) is built 
from a single one matrix \(H\). This is the 
case for the
\((1,0)\) and \((0,1)\) geometries on which we focus.
For a given matrix \(H\) the corresponding 
Dirac operator \(D\) describes one finite non-commutative geometry.
%While it is not relevant for the present paper how this is achieved
%let us just mention that this is achieved in analogy to reconstructing
%a Laplacian operator on a Riemannian manifold as a differential operator
%from its matrix representation in some (unknown) basis .
A toy model for quantum gravity is obtained  
as a weighted sum over geometries represented by Dirac operators \(D_\pm\).
Effectively,
the model reduces to a 
random-matrix ensemble
with a probability
measure 
\begin{equation} d\mu(H) = P(H) dH= \frac{1}{Z} 
e^{-N^2S(H) }dH
\end{equation}
where \(dH\) is the flat measure on the space of hermitian
matrices (with constraint \( \mathrm{tr}\, H=0\) in the \((0,1)\) case).
Random-matrix ensembles of this form 
belong to the more general form of unitarily invariant
ensembles that are central to many applications of random-matrix theory 
\cite{mehta_book, forrester, handbook}
and its applications \cite{QSOC}.
For the more general \((p,q)\) case \(H\)
stands for \(2^{p+q-1}\) matrices and \(dH\)
for the product of flat measures for each 
matrix (constrained to zero trace 
for anti-commuting parts). 
The normalization is given in terms of the partition sum
\begin{equation}
     Z = \int  e^{-N^2S(H)} dH
\end{equation}
Barrett and Glaser first considered a quadratic 
action 
\(S= \frac{1}{N^2}\ \mathrm{tr}\ D^2\)
and showed that standard results from random-matrix theory 
carry over to this model for large \(N\). 
For instance, in the one-matrix models 
\((1,0)\) and \((0,1)\) the eigenvalues of the 
underlying hermitian matrix \(H\) are distributed according to 
the Wigner semicircle law while the eigenvalues of \(D\)
are given as a convolution of the semicircle law. In 
other geometries that are 
described by two or more matrices the eigenvalues have limiting
distributions that follow different laws and that show soft gaps or 
additional peaks in the centre of the \(H\) spectrum.\\
In order to describe phase transitions or crossovers, Barrett and Glaser 
also introduced the one-parameter family 
\begin{equation}
  S_g(D) \equiv S_g(H)=\frac{1}{N^2} \mathrm{tr}\, D^4 + 
  \frac{g}{N^2} 
  \ \mathrm{tr}\, D^2
\end{equation}    
which they considered for general 
\((p,q)\) geometries.
For the \((1,0)\) and \((0,1)\) geometries these
can also be written in the form
\begin{align}
  S^{\pm}_g(H)=
  & \frac{2}{N}  \left( \mathrm{tr}\, H^4 + g\ \mathrm{tr}\, H^2 \right)
  \pm \frac{2g}{N^2} \left(\mathrm{tr}\, H \right)^2\pm \frac{8}{N^2}
  \ \mathrm{tr}\, H \ 
  \mathrm{tr}\, H^3 + \frac{6}{N^2} \left( \mathrm{tr}\, H^2\right)^2\nonumber\\
  =&
  \begin{cases}
  \frac{2}{N}  \left( \mathrm{tr}\, H^4 + g\ \mathrm{tr}\, H^2 \right)
  + \frac{2g}{N^2} \left(\mathrm{tr}\, H \right)^2 + \frac{8}{N^2}
  \ \mathrm{tr}\, H \ 
  \mathrm{tr}\, H^3 + \frac{6}{N^2} \left( \mathrm{tr}\, H^2\right)^2
  &\text{for \((1,0)\)},
  \\
  \frac{2}{N}  \left( \mathrm{tr}\, H^4 + g\ \mathrm{tr}\, H^2 \right)
  + \frac{6}{N^2} \left( \mathrm{tr}\, H^2\right)^2&\text{for \((0,1)\).}
  \end{cases}
  \label{eq_action_H}
\end{align}
The dependence on the parameter \(g\) is chosen 
so that 
the action has the form of a generalized Mexican 
hat for \(g<0\).
Using Monte-Carlo simulations, 
Barrett and Glaser showed numerically that the 
spectra of
these models also have
limiting distributions and that there is a 
transition from a behaviour
dominated by operators \(\mathrm{tr}\, D^2 = O(1)\) for \(g>0\)
to \(\mathrm{tr}\, D^2 \propto -g\) for \(g\ll 0\)
with a crossover or phase transition (for
\(N \to \infty\)) at a critical
negative value of the coupling constant \(g_c<0\)
depending on the type \((p,q)\).
The critical value can be seen most clearly 
in the spectral measure which is supported on
one interval for \(g>g_c\) (this will be 
referred to as 1-cut) while 
for \(g<g_c\) the support splits into two 
separate intervals (2-cut).

Phase transitions and related 
spectral properties have been investigated in 
various \((p,q)\) geometries 
\cite{Glaser_2017,barrett2019,khalkhali2020,khalkhali2022,hessam2022-bootstrapping,hessam2022-review,azarfar2024random,barrett2024fermion,khalkhali2025large}
where the review \cite{hessam2022-review}
gives an overview of the work before 2022.
Related topics have been considered for a long time in
random-matrix theory.
Among them phase transitions in spectral densities that
are accompanied by a separation of the support 
in may intervals (known as multi-cut solutions) \cite{jurkiewicz,ambjorn1993matrix,akemann1996JPA,akemann1996NPB,Bonnet}. 
We use standard methods of random-matrix theory such as 
the description of eigenvalue spectra of 
random matrices as a Coulomb gas (see \cite{forrester}
and references for a general overview of these methods
in random-matrix theory). In this framework the
Riemann-Hilbert approach is a standard 
method -- this is described with mathematical rigor 
in \cite{deift2000}.
In our case we have 
to incorporate additional interaction terms due 
to the appearance of squares of traces in the 
action -- such models have appeared in random-matrix theory before, e.g. 
\cite{iso1997}.
 
Our contribution is a direct response to the 
work of Khalkhali and 
Pagliaroli \cite{khalkhali2020} on the asymptotic
density of states 
as the dimension grows to infinity. 
They consider the \((1,0)\) and \((0,1)\) 
models and derived analytical expressions for 
the spectral measure and the critical 
value of the coupling constant using 
a Riemann-Hilbert approach in the Coulomb-gas 
description \cite{deift2000}. 
As part of that they find explicit expressions for
1-cut spectral measures for \(g>g_{\mathrm{cr}}\)  and
2-cut measure for \(g<g_{\mathrm{cr}}\).
Comparing their results to the 
numerical results of Barrett and Glaser shows 
that for \((0,1)\) there is strong 
qualitative agreement and somewhat weaker
quantitative agreement (Khalkhali and 
Pagliaroli suggest that this 
is due to the relatively 
small dimensions in the numerical simulations). 
However, in the \((1,0)\) 
the difference is strong even on a 
qualitative level as there are signatures of
a symmetry breaking \(H\to -H\)
(which is an obvious symmetry of \(S_g(H)\))
in the numerical simulations 
which are absent in the analytical approach.
Indeed Khalkhali and 
Pagliaroli explicitly \emph{assume} that a 
well-known existence and uniqueness theorem 
\cite{BPS, Johansson, deift2000} for
spectral measures for random matrix ensembles 
of the form
\(d \mu_V(H) = e^{- N\,  \mathrm{tr}\, V(H)} dH \)
can be extended to the present case.
The uniqueness part of that assumption 
basically rules out symmetry breaking.
As the numerical simulations were 
done at relatively small
dimension this raises the question whether 
symmetry breaking persists in the 
limit \(N \to \infty\) or the explicit solutions
by Khalkhali and Pagliaroli eventually take over
when \(N\) is sufficiently large.
As the title of this work suggests our 
work demonstrates that symmetry breaking persists.
We do this by repeating the derivation of
Khalkhali and Pagliaroli without
any symmetry assumptions.
We have also found a few calculational errors
in their work that we correct.
For the \((0,1)\) case our approach 
strengthens the argument by Khalkhali and 
Pagliaroli by giving a clearer reason why, in 
this case, the uniqueness theorem applies.
In addition, the correction of the calculational 
errors leads to a much better 
quantitative agreement for the critical coupling 
constant \(g_{-,c}\) and the spectral measure.
In the \((1,0)\) case we 
will show that the asymptotic 
spectral measure jumps
abruptly from a symmetric 1-cut solution 
for \(g>g_{+,c}\) to a broken symmetry 2-cut 
solution for \(g< g_{-,c}\). For 
this one needs to distinguish the 
asymptotic spectral measures from the mean 
density of states, as the latter is 
always symmetric.
We give the explicit form of the broken-symmetry
2-cut solutions which contains a number 
of parameters that depend on the coupling constant
\(g\) through a number of nonlinear 
implicit equations (which can be 
solved numerically by standard Newton-Raphson methods).

The different behaviour of the two models can already be seen
on a very basic level by considering the order parameter
\(M = \frac{1}{N} \mathrm{tr} H\) and writing
\(H= M + H'\) where \(H'\) is traceless.
In that case \(S^{-}_g (H)= S^{-}_g(H')\) while
\(S^{+}_g(H)= 16 M^4 + 4gM^2  + S^{+}_g(H') + F(M,H')\).
In the \((0,1)\) case \(S^{-}_g\) does not depend on this order
order parameter. In the \((1,0)\) case  \(S^{+}_g\)
contains the term \(16 M^4 + 4gM^2\) which (on its own) is equivalent
to the magnetization of a ferromagnet which is a paradigm of a phase 
transition that is accompanied by symmetry breaking. 
This simple consideration can at most give an indication.
The full answer is given in this paper below (together with the
explicit expressions for spectral measures and critical value of the coupling constant for a phase transition.

% gcrit= -3.187
\section{The Coulomb gas approach to spectral measures and 
the density of states
}

\subsection{The Coulomb gas description}

The family of random matrix ensembles described by the 
action \eqref{eq_action_H}
belongs to the class of unitary
invariant ensembles where
\(d\mu_{\pm, g}(H) = d\mu_{\pm, g} (UHU^\dagger)\) for any unitary matrix \(U\)
(like the Gaussian Unitary Ensemble GUE)).
In this case
spectra are statistically independent of eigenvectors. 
Writing
\( H= U \Lambda U^\dagger\)
in terms of its eigenvalues
\(\Lambda=\mathrm{diag}(\lambda_1,\dots,\lambda_N)\)
and a diagonalizing unitary matrix,
one has \(d\mu_{\pm, g}(H) = d\mu_{\mathrm{Haar}}(U) d\tilde{\mu}_{\pm,g}(\Lambda)\)
where \(d\mu_{\mathrm{Haar}}(U)\)
is the Haar measure on the coset space \(U(N)/U(1)^N\), and
\begin{equation}
  d \tilde{\mu}_{\pm, g}(\Lambda)= \frac{1}{\tilde{Z}(g)} 
  \prod_{i< j}
  \left|\lambda_i-\lambda_j\right|^2 e^{-N^2 S_{\pm, g}
  (\Lambda)} d\Lambda= \frac{1}{\tilde{Z}(g)} 
  e^{-N^2 \mathcal{E}_{\pm, g}
  (\Lambda)} d\Lambda
  \label{eq_joint_probability_eigenvalues}
\end{equation}
is the joint probability measure of the eigenvalues.
Note that the joint probability measure is symmetric under any exchange of
eigenvalues.
Here
\(d\Lambda= d\lambda_1 d\lambda_2 \dots d\lambda_N\)
is just the flat measure on \(\mathbb{R}^N\)
(in the \((0,1)\) we add an additional factor 
\(\delta(\sum_{k=1}^N \lambda_k)\) to \(d\Lambda\)).
The factor \(\prod_{i< j}
  \left|\lambda_i-\lambda_j\right|^2\) is a Vandermonde
determinant that enters as part of the Jacobean 
of the coordinate transformation.
This factor has been absorbed into the exponent in the 
expression on the right.
The latter is the partition sum for a 
one-dimensional Coulomb gas with \(N\) particles
at positions \(\lambda_i\) where
\(N^2 \mathcal{E}_{\pm, g}\) is the energy density of the gas.
Let us introduce the normalized density of the
Coulomb gas
\begin{equation}
    \rho_{\Lambda}(\lambda)= \frac{1}{N} \sum_{k=1}^N \delta(\lambda-\lambda_k)
\end{equation}
which is also known as the \emph{density of states}.
The mean density of states averaged over the ensemble is
just the marginal probability density of one eigenvalue
\begin{equation}
   \mathbb{E}\left[ \rho_{\Lambda}(\lambda) \right]=
   \int  \rho_{\Lambda}(\lambda) d\tilde{\mu}_{\pm,g}(\Lambda)=
   \mathbb{E}\left[ \delta(\lambda-\lambda_1) \right]\ .
\end{equation}
The behaviour of the density of states as \(N \to \infty\) is at the centre
of this work. We can now write
\begin{equation}
\begin{split}
    \mathcal{E}_{\pm, g}(\Lambda)
    \equiv \mathcal{E}_{\pm, g}\left[ \rho_\Lambda(\lambda)\right]
    =& \int_{\mathbb{R}} V_g(\lambda) \rho_\Lambda(\lambda)\ d\lambda
    + \int_{\mathbb{R}^2} U_{\pm,g}(\lambda,\lambda') \rho_\Lambda(\lambda)
    \rho_\Lambda(\lambda')\ d\lambda d\lambda'\\
    &\; + \int_{\mathbb{R}^2: \lambda\neq \lambda'} 
    \log\left(\frac{1}{|\lambda-\lambda'|}\right) 
    \rho_\Lambda(\lambda) \rho_\Lambda(\lambda')\ d\lambda d\lambda'
    \end{split}
    \label{eq_energy_functional}
\end{equation}
with
\begin{equation}
    V_g(\lambda) =2\lambda^4+2g \lambda^2
\end{equation}
and
\begin{equation}
    U_{\pm,g}(\lambda,\lambda')
    =
    \begin{cases}
        2g\lambda \lambda'+4\lambda {\lambda'}^3+4\lambda^3\lambda'+6\lambda^2
        {\lambda'}^2& \text{for \((1,0)\)}\\
        6\lambda^2{\lambda'}^2& \text{for \((0,1)\).}
    \end{cases}
\end{equation}
In the last (logarithmic) term  of \eqref{eq_energy_functional} the double integral
over \((\lambda,\lambda')\in \mathbb{R}^2\) needs to exclude a strip 
\(|\lambda-\lambda'|<\epsilon\) for arbitrarily
small \(\epsilon >0\) in order to reproduce the Vandermonde determinant in 
\eqref{eq_joint_probability_eigenvalues}. In the following, we will consider 
\(\mathcal{E}\left[ \rho(\lambda)\right]\) as a functional over regular 
(continuous, piecewise differentiable) densities \(\rho(s)\) where this exclusion is not necessary.

\subsection{Some known random-matrix results for the asymptotic density of states}
\label{sec_rmtresults}

One of the seminal results of random-matrix theory for Gaussian
Wigner-Dyson
ensembles 
is the Wigner semicircle law \cite{wigner1955}.
Let us explain this briefly for the Gaussian unitary ensemble GUE and consider
a sequence of hermitian random matrices \(H_N\)
of dimension \(N\)
that are drawn from the Gaussian probability measure
\begin{equation}
   d\mu_{GUE}(H)= \frac{1}{Z_{GUE}} e^{-\frac{N}{2} \mathrm{tr} H^2} dH \ .
\end{equation}
Denote the corresponding eigenvalues \(\Lambda_N=(\lambda_{1,N},\dots,\lambda_{N,N})\).
The Wigner semicircle law then states that the corresponding density of states converges 
(in a weak sense) to a semicircle law
\begin{equation}
    \rho_{\Lambda_N}(\lambda) \to \rho_{\mathrm{sc}}(\lambda)= 
    \begin{cases}
        \frac{\sqrt{4-\lambda^2}}{2 \pi} & \text{if \(|\lambda| \le 2\),}\\
        0 & \text{if \( |\lambda| >2 \).}
    \end{cases} 
\end{equation}
Without going into full technical detail weak convergence here means that for 
any (sufficiently nice) function \(f(\lambda)\) one has 
\begin{equation}
    \mathbb{E}
    \left[
    \left(\int f(\lambda) 
    \left(\rho_{\Lambda_N}(\lambda) -\rho_{\mathrm{sc}}(\lambda)
    \right)d\lambda \right)^2
    \right]
    \to 0 \ 
\end{equation}
as \(N \to \infty\) where the expectation value is taken with respect to the GUE of dimension 
\(N\).
More simply put, for a large matrix \(H\) drawn at random from GUE 
the difference between 
\(\frac{1}{N} \sum_{k=1}^N f(\lambda_k)\) and 
\(\int f(\lambda)\rho_{\mathrm{sc}}(\lambda)d\lambda \) is negligibly small and vanishes as
\(N \to \infty\). The semicircle law is a strong statement as it implies that
the spectrum of one typical large GUE matrix follows a semicircle law.
In particularly, it implies that the mean density of states converges to the semicircle distribution
(again in the weak sense)
\begin{equation}
    \mathbb{E}
    \left[
    \rho_{\Lambda_N}(\lambda)
    \right] \to \rho_{\mathrm{sc}}(\lambda)\ .
\end{equation}
Mathematically rigorous results are available
for a more general class of (hermitian) ensembles 
with probability measures of the form
\begin{equation}
    d\mu_{W}(H) = \frac{1}{Z_W} e^{- N\, \mathrm{tr}\, W(H)}
\end{equation}
for a given function \(W(\lambda)\) \cite{deift2000}. 
Such ensembles have been considered in 
random-matrix theory in a lot of detail (see e.g.
\cite{ambjorn1993matrix, akemann1996JPA, akemann1996NPB, Bonnet})
Normalization requires
\(W(\lambda)\to \infty \) as \(\lambda \to \pm \infty\) though
technical conditions on the rate of growth are assumed.
We will focus on the simple case 
of a finite polynomial 
\begin{equation}
    W(\lambda)= \sum_{k=0}^{2M} w_k \lambda^{k}\
    \end{equation}
of even order \(2M\) with positive highest coefficient \(w_{2M}>0\).
GUE belongs to this class as the special case \(W_{GUE}(\lambda)= \frac{\lambda^2}{2}\).
For this class the density of states has a unique weak limit
\begin{equation}
    \rho_{\Lambda_N}(\lambda) \to \rho_{W}(\lambda)
\end{equation}
where the asymptotic spectral density \( \rho_{W}(\lambda)\)
is the minimizer of the free energy functional
\begin{equation}
    \mathcal{E}_{W}[\rho(\lambda)] =
    \int_{\mathbb{R}^2} \log\left(\frac{1}{\left|\lambda-\lambda'\right|}\right)
    \rho(\lambda)\rho(\lambda') d\lambda d\lambda' + \int_{\mathbb{R}} W(\lambda) 
    \rho(\lambda) d\lambda
    \label{eq_free_energy_W}
\end{equation}
over all non-negative densities \(\rho(\lambda)\ge 0\) that are normalized
\(\int_{\mathbb{R}} \rho(s) ds =1\).
It is therefore also referred to as the equilibrium measure.
The explicit form of the equilibrium measure can often be derived using
the Riemann-Hilbert approach. A common feature of these equilibrium measures (shared by
the semicircle law as a special case) is that they have a compact support that consists of
a finite number of intervals.

\subsection{Outline of the Riemann-Hilbert method}
\label{sec_RH_method}

For completeness and later use let us 
shortly summarize the main steps and ingredients of the Riemann-Hilbert
approach.
A detailed and rigorous exposition can be found in \cite{deift2000}.
We do not require rigour and full generality here and focus on the special cases relevant
later.
Let us thus assume that there is an 
equilibrium measure (non-negative and normalized)
\(\rho_{W}(\lambda)\)
that minimizes the free energy \eqref{eq_free_energy_W} and
its support, denoted as \(\Omega\), is a disjoint union of finitely many intervals
\begin{equation}
\Omega= \bigcup_{k=1}^K [a_k,b_k]\qquad
-\infty < a_1<b_1<a_2\dots <a_K<b_K<\infty. 
\end{equation}
The positive integer \(K\) is referred to as the number of cuts. 
The aim is to find an explicit expression
for \(\rho_{W}(\lambda)\) (including explicit values for the interval 
boundaries \(a_k\) and \(b_k\)).
Variation of the free energy implies that the equilibrium 
measure satisfies the conditions
\begin{subequations} \begin{align}
    2\int_{\Omega} \log\left(\frac{1}{|\lambda-\lambda'|}\right) 
    \rho_{W}(\lambda') d 
   \lambda'+ W(\lambda) &\ge \ell & \text{for \(\lambda \in \mathbb{R}\),} 
 \label{eq_cond_a}\\
    2\int_{\Omega} 
    \log\left(\frac{1}{|\lambda-\lambda'|}\right) 
    \rho_{W}(\lambda') 
     d \lambda'
     + W(\lambda) &= \ell & 
     \text{for \(\lambda \in \Omega \).%
     }\label{eq_cond_b}
    \end{align}\label{eq_cond}\end{subequations}%
Here, \(\ell\in \mathbb{R}\) is a Lagrange multiplier that ensures the 
normalization 
constraint. We will not need the value that \(\ell\) takes, 
we just need to assume
that this constant exists. Note that in the equality \eqref{eq_cond_b} the 
left hand 
depends explicitly on \(\lambda \in \Omega\) while the right side is a 
constant.\\
The Riemann-Hilbert approach is a standard method to find the equilibrium 
measure
\(\rho_{W}(\lambda)\) explicitly from the conditions \eqref{eq_cond}.
Before proceeding let us introduce the Borel transform
of \(\rho_{W}(\lambda)\)
\begin{equation}
    G_{\rho_{W}}(z)=\frac{i}{\pi} 
    \int_{\Omega} \frac{\rho_W(\lambda)}{z-\lambda} d\lambda
    \quad \text{for}\ z \in \mathbb{C}\setminus \Omega
 \end{equation}
and the Hilbert transform 
of \(\rho_{W}(\lambda)\)
\begin{equation}
    H_{\rho_{W}}(x)= \frac{1}{\pi} \mathrm{PV} \int_\Omega 
    \frac{\rho_W(\lambda)}{x-\lambda} 
    d\lambda \equiv \lim_{\epsilon \to 0}
    \frac{1}{\pi}\int_\Omega \frac{(x-\lambda)
    \rho_W(\lambda)}{(x-\lambda)^2 + 
    \epsilon^2} 
    \quad \text{for}\ x \in \mathbb{R}.
\end{equation}
The two functions are obviously closely related
as
\begin{subequations}
    \begin{align}
    G_{\rho_{W}}(x) &=& i    H_{\rho_{W}}(x) 
    &&&\text{for}\ x \in \mathbb{R}\setminus \Omega,\\
    \quad\lim_{\epsilon \to 0^+} G_{\rho_{W}}(x\pm i \epsilon)
    &=& \pm \rho_W(x) +
    i H_{\rho_{W}}(x) &\quad \qquad&&\text{for}\ x \in \Omega. \qquad
    \end{align}
\end{subequations}
Let us introduce the moments of the equilibrium measure 
\begin{equation}
    m_n = \int_\Omega  \lambda^n \rho_W(\lambda) d\lambda\ .
    \label{eq_def_moments}
\end{equation}
Normalization of the measure implies \(m_0=1\).
The asymptotic expansion
\begin{equation}
    G_{\rho_{W}}(z)=\frac{i}{\pi}\sum_{n=0}^\infty
    z^{-(n+1)} m_n \qquad \text{as}\ |z|\to \infty
\end{equation}
shows that the Borel transform (and hence the Hilbert transform)
are moment generating functions for the equilibrium measure.
Taking the derivative of the condition \eqref{eq_cond_b} 
one obtains
\begin{equation}
    2\pi H_{\rho_{W}}(x) = \frac{dW}{dx}(x)\equiv W'(x) 
    \qquad \text{for}\ x\in \Omega.
\end{equation}
The problem of finding \(G_{\rho_{W}}(z)\) given
\(H_{\rho_{W}}(x)\) on \(x \in \Omega\) can be cast into a standard 
scalar Riemann-Hilbert problem which can be solved using the 
Plemelj formula (see \cite{deift2000} for details)
and this leads to the expression
\begin{align}
    G_{\rho_{W}}(z)=&\frac{\sqrt{q(z)}}{2\pi^2}
    \int_{\Omega} \frac{W'(x)}{\left(\sqrt{q(x)}\right)_+ } 
    \frac{dx}{x-z}\nonumber\\
    =& \frac{i}{2\pi} W'(z) +\frac{\sqrt{q(z)}}{4\pi^2}
    \oint_{\mathcal{C}}
    \frac{W'(z')}{\sqrt{q(z')}}\frac{dz'}{z'-z}.
    \label{eq_Borel_Plemelj}
\end{align}
Here
\begin{equation}
    q(z)= \prod_{k=1}^K (z-b_k)(z-a_k)
\end{equation}
is a polynomial of order \(2K\) and the branch cuts of
\(\sqrt{q(z)}\) are chosen to coincide with the support \(\Omega\)
such that \(\sqrt{q(z)} \sim z^K\) as \(|z|\to \infty \),
and we denote
\(\left(\sqrt{q(x)}\right)_+\equiv \lim_{\epsilon\to 0}
\sqrt{q(x+i\epsilon)}
\) for \(x\in \Omega\). The closed contour \(\mathcal{C}\) in the final expression
of \eqref{eq_Borel_Plemelj}
encircles the support \(\Omega\) and the point \(z\) clockwise.
The contour integral can be calculated using the asymptotic expansion of
the integrand 
%%Change integrant -> integrand
for \(|z|\to \infty\) (evaluating the residue 
at infinity). \\
The expression \eqref{eq_Borel_Plemelj} still depends explicitly on the
boundaries \(a_k\) and \(b_k\). The uniqueness of the equilibrium 
measure implies that these are completely determined from the polynomial
\(W(\lambda)\). So we still miss \(2K\) conditions.
%% Change 'The first set of conditions can be found by requiring that
%%\eqref{eq_Borel_Plemelj} satisfies the consistency requirements
%% for \(G_{\rho_W}(z)\). ->  following phrase
The first set of conditions can be found by requiring that the 
asymptotics for large \(|z|)\) 
are consistent with the definition of \(G_{\rho_W}(z)\).
The latter implies \(G_{\rho_W}(z)\sim \frac{i}{\pi z} + O(z^{-2})\)
for large \(z\). One can show that the right-hand side
of \eqref{eq_Borel_Plemelj} has an expansion
%%change: \sum_{n=0^\infty} -> \sum_{n=0}^\infty
\begin{equation}
    G_{\rho_W}(z)= \sum_{n=0}^\infty g_n z^{K-1-n}
\end{equation}
with coefficients that depend on the boundaries.
This gives \(K+1\) conditions
\begin{equation}
    g_n=0\quad \text{for}\ 0\le n\le K-1\quad \text{and} \quad
    g_k= \frac{i}{\pi}.
\end{equation}
The remaining \(K-1\) conditions are obtained by requiring that
the Lagrange multiplier \(\ell\) in \eqref{eq_cond_b} takes the same
value in each interval. This leads to \(K-1\) conditions \cite{deift2000}
\begin{equation}
\int_{b_k}^{a_{k+1}} \left(G_{\rho_W}(x) - i \frac{W'(x)}{2\pi} \right)
dx=0 \quad \text{for}\ 1\le k \le K-1.
\end{equation}
In the physical Coulomb gas description this
follows from the requirement that, in equilibrium,
the chemical potential
of the gas be the same in all intervals
\cite{jurkiewicz}.

In Appendix~\ref{app_Wlambda} we give more explicit results
for the cases that are relevant for discussing the random-matrix 
ensembles underling for \((1,0)\) and \((0,1)\) geometries.

\section{The asymptotic spectral measures for \((0,1)\) and \((1,0)\) geometries}

\subsection{Applying the Riemann-Hilbert approach in the present setting}

Let us now return to the main topic and discuss the random-matrix models 
for the \((0,1)\) and \((1,0)\) random geometries. We have already introduced 
the Coulomb gas description. Existence and uniqueness theorems are not available 
for this setting because \(S^{\pm}_g(H)\) contains products of traces.
Continuing along the same lines as Khalkhali and Pagliaroli \cite{khalkhali2020} let us consider 
the free energy functional \(\mathcal{E}_{\pm,g}[\rho(\lambda)]\) given
in \eqref{eq_energy_functional} and use the Riemann-Hilbert approach 
to find the equilibrium measures 
\(\rho_{\pm,g}(\lambda)\) which minimize the free energy functional 
for these cases. We continue to denote the support of \(\rho_{\pm,g}(\lambda)\)
by \(\Omega \subset \mathbb{R}\).
Constrained variation of the functional here leads to 
\begin{subequations}
    \begin{align}
        2\int_{\Omega}
        \log
        \left(
            \frac{1}{|\lambda-\lambda'|} 
        \right)
        \rho_{\pm,g} d\lambda' 
        +
        2 \int_{\Omega}
        U_{\pm,g}(\lambda,\lambda')\rho_{\pm,g} (\lambda') d\lambda' 
        + V_{g}(\lambda) &\ge\ell && \text{for}\ \lambda \in 
        \mathbb{R},\\
        2\int_{\Omega}
        \log
        \left(
            \frac{1}{|\lambda-\lambda'|} 
        \right)
        \rho_{\pm,g} d\lambda' 
        +
        2 \int_{\Omega}
        U_{\pm,g}(\lambda,\lambda')\rho_{\pm,g} (\lambda')  d\lambda' 
        + V_{g}(\lambda) &= \ell && \text{for}\ \lambda \in \Omega,\ 
        \label{eq_cond_U}
    \end{align}
\end{subequations}
We mention in passing that Khalkhali and Pagliaroli have some minor
calculational errors in the expressions above which we corrected. Apart from 
the corrections, we  
continue with their approach. 
Taking the derivative of the second condition with respect to \(\lambda\)
gives
\begin{subequations}
    \begin{equation}
        2\pi H_{\rho_{\pm,g}}(\lambda) + 
        2 \int_{\Omega}
            U'_{\pm,g}(\lambda,\lambda')\rho_{\pm,g} (\lambda')  d\lambda' 
            + V'_{g}(\lambda)= 0 \qquad \text{for}\ \lambda \in \Omega
            \label{eq_cond_derivative}
    \end{equation}
    where
        \begin{align}
        V'_g(\lambda)=& 8 \lambda^3 + 4 g \lambda, & \text{and}\\
        U'_{\pm,g}(\lambda, \lambda')=&
        \begin{cases}
            12 \lambda \lambda'^2+2g\lambda' +4 \lambda'^3+12 \lambda^2 \lambda'
            & \text{for \((1,0)\), and}\\
            12 \lambda \lambda'^2& \text{for \((0,1)\).}
        \end{cases}
    \end{align}
\end{subequations}
In order to use the Riemann-Hilbert method
as laid out in Section~\ref{sec_RH_method} the next step is to (formally) integrate 
the term containing \(U'_{\pm,g}(\lambda,\lambda')\) in \eqref{eq_cond_derivative}. 
Writing the latter as
\begin{equation}
    2\pi H_{\rho_{\pm,g}}(\lambda) + 
        W'_{\pm,g}(\lambda) = 0
        \label{eq_cond_eff}
\end{equation}
with
\begin{equation}
    W'_{\pm,g}(\lambda) =
    \begin{cases}
        8\lambda^3 +24m_1 \lambda^2 +4(6m_2+g)\lambda+ 8m_3+4gm_1
        & \text{for \((1,0)\), and}\\
        8\lambda^3 +4(6m_2+g)\lambda
        & \text{for \((0,1)\).}
    \end{cases}
    \label{eq_W_eff}
\end{equation}
The coefficients \(m_n\) are the moments of the equilibrium measure as defined
in \eqref{eq_def_moments}.
We refer to the 4th-order polynomial \( W_{\pm,g}(\lambda) \)
%%change: deleted 'with \(W_{\pm,g}(\lambda)\) 
%% and derivative given by \eqref{eq_W_eff}'
as the effective potential.
Using the Riemann-Hilbert approach, one may find the equilibrium measure 
and its
Borel transform for 
\begin{align}
W'(\lambda)=&4w_4\lambda^3 + 3 w_3\lambda^2 + 2 w_2 \lambda + w_1\ .
\end{align}
This is done in detail in Appendix~\ref{app_Wlambda}.
For each of these solutions one may find all moments from the asymptotic expansion of the
Borel transform.
By requiring consistency through
\begin{align}
w_1&=
\begin{cases}
8m_3+4gm_1 & \text{for \((1,0)\)}\\
0 & \text{for \((0,1)\)}
\end{cases}&
w_3&=
\begin{cases}
8 m_1 & \text{for \((1,0)\)}\\
0 & \text{for \((0,1)\)}
\end{cases}\nonumber \\
    w_2&=2(6m_2+g)
    &
    w_4&=2
\end{align}
one then obtains the equilibrium measures of the 
random-matrix models for
\((1,0)\)
and \((0,1)\) geometries.
The condition \(w_4=2\) is trivial. 
In the \((1,0)\) the other three conditions are non-trivial.
In the \((0,1)\) case there is only one non-trivial condition
and \(w_1=w_3=0\) further simplify problem.
In appendix~\ref{app_Wlambda} we have derived detailed equations for
unconstrained values of the coefficients \(w_n\). 
Note that in either case the additional (non-linear) conditions may have
more than one solution. Whenever there is more than one solution comparing the free energies and
choosing the smallest value will give the relevant equilibrium measure. 

\subsection{Equilibrium measures, phase transitions, and symmetry breaking}

Let us now consider the models for the two types of geometries and discuss 
the main features of the model as a function of the coupling constant \(g\).
We refer to the Appendices~\ref{app_Wlambda} and \ref{app_2cut_conditions}
for all technical details.
Khalkhali and Pagliaroli have assumed that, due to 
the symmetry of the action \(S_g(H)=S_g(-H)\), 
the equilibrium density is also symmetric
\(\rho_{\pm,g}(\lambda)=\rho_{\pm,g}(-\lambda)\).
They refer
to the existence and uniqueness proof for random matrix ensembles with
probability measure \(\frac{1}{Z}e^{- N\, \mathrm{tr}\, W(H)} dH\)
and \textit{assume} that it generalizes to 
the ensembles considered here. 
%The 
%symmetric action \(S_g(H)=S_g(-H)\) of these models 
%and the assumed uniqueness then imply symmetric equilibrium densities.
They are aware that the existing mathematical proof
does not extend to the present case.
The assumption is basically made to get a start on the problem -- and is
partly backed by numerical results.
We will show that their argument can be strengthened in one case 
but fails in the other.

\subsubsection{The case of \((0,1)\) geometries}

We start with the simpler case of \((0,1)\) geometries.
The effective potential in this case is always
symmetric \(W_{-,g}(\lambda)=W_{-,g}(-\lambda)\).
The variational conditions in the \((0,1)\)
case are equivalent to random-matrix models where 
uniqueness hold. 
One just has to add as a constraint that the parameter \(m_2\)
that appears in  \(W_{-,g}(\lambda)\) is equal
to the second moment of the equilibrium density.
As uniqueness holds for any 
%%change value given -> given value
given value of \(m_2\),
the support and density are always symmetric.
The additional constraint is nonlinear in nature and 
may, in principle, lead
to more than one solution for the equilibrium density 
\(\rho_{-,g}(\lambda)\) (in which case the relevant solution is the 
one with smaller free energy).\\
For \(g>0\) the potential \(V_{g}(\lambda)\) and the effective potential 
\(W_{-,g}(\lambda)\) 
have a single well and one expects 1-cut solutions
with symmetric support \(\Omega= [-b,b]\) for \(b>0\).
The 1-cut equations given in Appendix~\ref{app_Wlambda} then 
simplify to
\begin{subequations}
\begin{align}
    G_{-,g}^{\text{1-cut}}(z)=&
    -\frac{\left(\frac{4}{b^2}-6b^2\right) z+ 8 z^3}{2 \pi i}+
    \frac{\sqrt{z^2-b^2}\left(
        \frac{4}{b^2}-2b^2 +8z^2
    \right)}{2\pi i}\\
    \rho_{-,g}^{\text{1-cut}}(\lambda)=&
    \begin{cases}
        \frac{\sqrt{b^2- \lambda^2}
        }{\pi}\left(\frac{2}{b^2}-b^2
        +4\lambda ^2
    \right)& \text{for \(|\lambda|\le b\)}\\
    0 & \text{for \(|\lambda|>b\)}
    \end{cases}
    \label{eq_1cutdensity}
    \\ 
    g=&\frac{1}{b^2}-3b^2-\frac{3}{4}b^6  \\  
    m_2=& \frac{b^2}{4}+\frac{b^6}{8}
\end{align}
\end{subequations}
Note that \eqref{eq_1cutdensity} gives a well defined (non-negative) 
density only if 
%%change < -> \le and > -> \ge
\(b^4 \le 2\) which implies \(g\ge  - 4 \sqrt{2}\).
This extends the existence of these solutions to some negative values
of the coupling constant.
For \(g\to \infty\) 
the density approaches a rescaled semi-circle law.
\\
Next we consider the 2-cut solutions (see Appendix~\ref{app_Wlambda}
for details).
Again we can assume a symmetric support
\(\Omega =[-b,a] \cup [a,b]\) with \(0<a<b\).
These simplify to
\begin{subequations}
\begin{align}
    G_{-,g}^{\text{2-cut}}(z)=&
    -\frac{
    2(6m_2+g)z+ 8 z^3}{2 \pi i}+ \frac{ 8z
    }{2\pi i}\sqrt{(z^2-b^2)(z^2-a^2)}
    \\
    \rho_{-,g}^{\text{2-cut}}(\lambda)=&
    \begin{cases}
    \frac{4}{\pi}|\lambda|
    \sqrt{(b^2- \lambda^2)(\lambda^2-a^2)}
        & \text{for \(\lambda\in \Omega\),}\\
    0 & \text{else,}
    \end{cases}
    \label{eq_2cutdensity}
    \\ 
    b= \frac{\sqrt{-g+4\sqrt{2}}}{2\sqrt{2}}, &
    \qquad  a= \frac{\sqrt{-g-4\sqrt{2}}}{2\sqrt{2}} \quad \text{and} \quad 
    m_2=-\frac{g}{8}.
\end{align}
\end{subequations}
This 2-cut solution is valid only for \(g\le - 4\sqrt{2}\).

\begin{figure}
\includegraphics[width=0.75\textwidth]{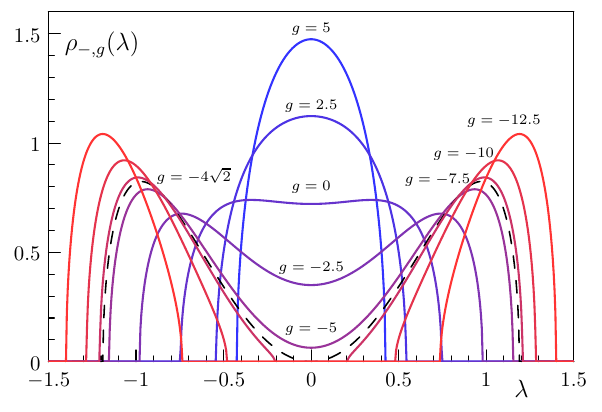}
\caption{\label{fig_01_rho} Equilibrium density (equivalent to 
expected density of states)
for the \((0,1)\) case for various values of the coupling constant \(g\). 
The dashed black curve is the density at the critical value \(g=-4\sqrt{2}\).}
\end{figure}
\begin{figure}
\includegraphics[width=0.49\textwidth]{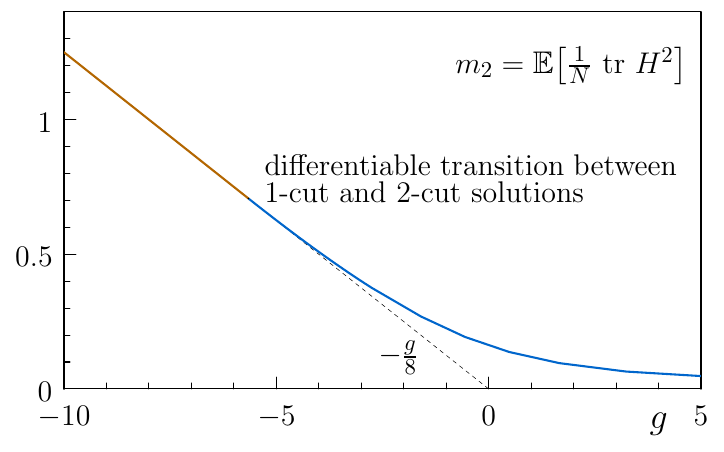}
\includegraphics[width=0.49\textwidth]{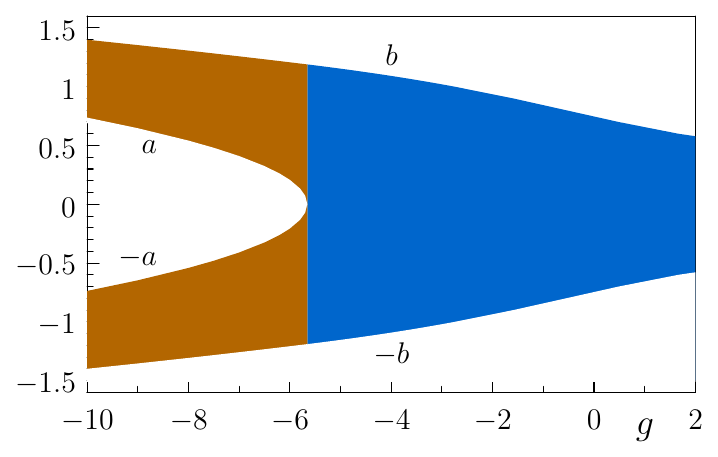}
\caption{\label{fig_01_m2} 
Left: Dependence of the
second moment
\(m_2\) 
on the coupling constant 
\(g\)
for the
\((0,1)\) case
(blue curve: dependence for \(g>-4\sqrt{2}\);
orange curve:
dependence for \(g<-4\sqrt{2}\);
dashed black curve: 
linear behavior \(-\frac{g}{8}\) which coincides with the orange curve
for \(g<-4\sqrt{2}\).\\
Right:
Support of the 
equilibrium density
for the
\((0,1)\) case. 
For \(g>-4\sqrt{2}\) the interval \([-b,b]\)
for the corresponding 1-cut 
solutions is drawn in blue.
%%Change \(g>-4\sqrt{2}\) -> \(g<-4\sqrt{2}\)
For \(g<-4\sqrt{2}\) the two symmetric intervals \([-b,-a]\)
and \([a,b]\) for the corresponding 
2-cut solutions are drawn in orange.
}
\end{figure}

%\begin{figure}
%\includegraphics[width=0.8\textwidth]{support_01.pdf}
%\caption{\label{fig_01_support} Support }
%\end{figure}

We have thus found a unique solution for any value of the coupling constant
\(g\) with a critical value \(g_{-,\mathrm{cr}}=-4\sqrt{2}\) as the
boundary between
the 2-cut and 1-cut symmetric solutions.
Up to corrections of calculation 
errors, the 1-cut and 2-cut solutions above (with the critical 
coupling constant separating the two) 
coincide with the findings of Khalkhali and Pagliaroli.
At the critical value, the two solutions are identical
\begin{equation}
    \rho_{-,-4\sqrt{2}}^{\mathrm{1-cut}}(\lambda)=
    \rho_{-,-4\sqrt{2}}^{\mathrm{2-cut}}(\lambda)=
    \frac{4 \lambda^2}{\pi}\sqrt{\sqrt{2}-\lambda^2} \ .
\end{equation}
The uniqueness of the solution implies that the equilibrium density is
equal to the expected density of states
\(\rho_{-,g}(\lambda)= \mathbb{E}\left[\rho_{\Lambda}(\lambda)\right]\)
as \(N\to \infty\).
In Figure~\ref{fig_01_rho} the equilibrium density is 
plotted for some values
of the coupling constant.
%%change \( g\gg 0\) by lof large positive values \(g>0\)\)
For large positive values \(g > 0\) the density has a single peak with 
maximum value at \(\lambda=0\).
Decreasing the coupling constant close to \(g=0\), the density starts 
to develop 
a symmetric double peak. The latter becomes more pronounced 
for negative coupling constants
until one reaches the critical coupling constant where the 
density becomes zero at the 
center
\(\lambda=0\) and the 1-cut solution is replaced by a 2-cut solution. 
%%change \(g\ll 4 \sqrt{2}\) If the coupling constant \(g\) 
%% is much smaller than the critical value \(4 \sqrt{2}\) THEN
If the coupling constant \(g\) 
is much smaller than the critical value \(4 \sqrt{2}\) then
the two intervals of the support are
separated by a growing interval where the density vanishes identically
(see right graph in Figure~\ref{fig_01_m2} which shows how the support
changes with the coupling constant).\\
The left graph in Figure~\ref{fig_01_m2} shows how
\(m_2\equiv \frac{1}{N}\mathbb{E}[\mathrm{tr}\ H^2]\) depends on
the coupling constant. Note that the crossover from the 1-cut to the 2-cut
solution at \(g=-4\sqrt{2}\) is continuous with a continuous 
first derivative. A direct calculation shows that higher derivatives 
are not continuous. As a result, this crossover is formally a third-order 
phase transition.

\subsubsection{The case of \((1,0)\) geometries}

In the \((1,0)\) geometries, the effective potential \eqref{eq_W_eff}
is generally not symmetric. 
In Appendix~\ref{app_Wlambda} we give the general equilibrium densities
and their Borel transforms for a 4th-order polynomial potential.
For each of these solutions the boundary of the support
are solutions to a set of simultaneous non-linear equations
that follow from consistency requirements of the 
asymptotic expansion of Borel transform.\\
For 1-cut solutions there are two boundary values and two nonlinear
equations that need to be satisfied (see Appendix~\ref{app_Wlambda} 
and \ref{app_2cut_conditions} for 
explicit formulas). For 2-cut solutions we have four boundary values and 
four nonlinear equations.
In the present setting the 
coefficients of the effective potential \eqref{eq_W_eff}
contain the three moments \(m_k\) (\(k=1,2,3\)).
This adds three consistency conditions
that can 
be derived from the asymptotic expansion of the Borel transform
(see Appendix~\ref{app_2cut_conditions} for explicit formulas).
Together, we have \(5\)
nonlinear conditions for \(5\) parameters in
the 1-cut case and \(7\) non-linear conditions for \(7\) parameters
in the 2-cut case. 
For a given value of the coupling constant there may, in principle, be
many solutions to the nonlinear set of equations
-- and we 
are interested in the one with minimal free energy.\\
\begin{figure}
\includegraphics[width=0.75\textwidth]{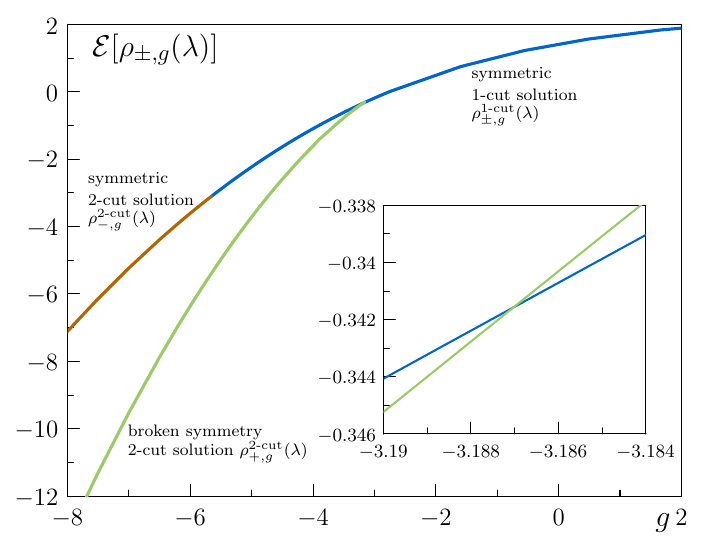}
\caption{\label{fig_free_energy} Free energy of various candidates for the equilibrium density of states in \((1,0)\) geometries.\\
Blue: the explicitly known symmetric 1-cut solution.\\
Orange: the explicitly known symmetric 2-cut solution.\\
Green: the numerically found broken symmetry 2-cut solution.\\
In the \((0,1)\) case only the symmetric
solutions (orange and blue) exist and they match at 
\(g_{-, \mathrm{cr}}=-4\sqrt{2}\). 
In the \((1,0)\) case the additional 
broken-symmetry solution minimizes the
free energy for \(g < g_{+,\mathrm{cr}}\).
The inset graph shows where the 
free energies of the symmetric 1-cut solution and the broken symmetry 2-cut solutions cross (the
broken symmetry solution exists beyond the crossing which cannot be seen in the large graph due to the width of the curves which hides the bifurcation point).
The critical value can be read off as 
\(g_{+,\mathrm{cr}}\approx -3.187\).
}
\end{figure}
The explicit conditions can be found in the appendices:
in Appendix~\ref{app_Wlambda} we derive the general form
of the equilibrium density for  arbitrary 4th order polynomial
potentials including the 
nonlinear conditions on the boundary of the support.
In Appendix~\ref{app_2cut_conditions} the additional 
conditions on the moments are added and the full set
of simultaneous non-linear equations is simplified as much as possible.\\
The equations simplify if one \emph{assumes} that 
the equilibrium density is a symmetric function.
In that case \(m_1=m_3=0\) and the effective 
potential is exactly the same as in the \((0,1)\)
case discussed above. The symmetric 1-cut and 2-cut
solutions solve the nonlinear conditions in the \((1,0)\)
case as well, \(\rho_{+,g}^{\mathrm{1-cut}}(\lambda)\equiv
\rho_{-,g}^{\mathrm{1-cut}}(\lambda)
\) and \(\rho_{+,g}^{\mathrm{2-cut}}(\lambda)\equiv
\rho_{-,g}^{\mathrm{2-cut}}(\lambda)
\). In their work
Khalkhali and Pagliaroli continue with that assumption
just as they did in the \((0,1)\) case. They conjecture that the 
asymptotic spectral densities for the two geometries have the same
behaviour (on the level of underlying matrices \(H\), the spectra
of the corresponding Dirac spectra would still behave differently).
While we could strengthen their 
argument in the \((0,1)\)
case where the effective potential was always symmetric, this is not the case in the 
\((1,0)\) case that we consider now.
Here, the parameter space is larger and the
corresponding effective potentials are in general not symmetric --
so there is no strong reason for excluding solutions with broken symmetry.\\
Finding such solutions is generally 
challenging. 
Assuming symmetry, as done by 
Khalkhali and Pagliaroli,
they reduce to the same equations as in the
\((0,1)\) case with the same
explicit solutions. 
However,  the symmetric solutions
are only seen in Monte-Carlo simulations 
when the coupling constant
is well above the transition. Since 
simulations are not done
at finite dimensions
Khalkhali and Pagliaroli
suggested that deviations 
from Monte-Carlo simulations may
disappear as \(N \to \infty\).\\
\begin{figure}
\includegraphics[width=0.75\textwidth]{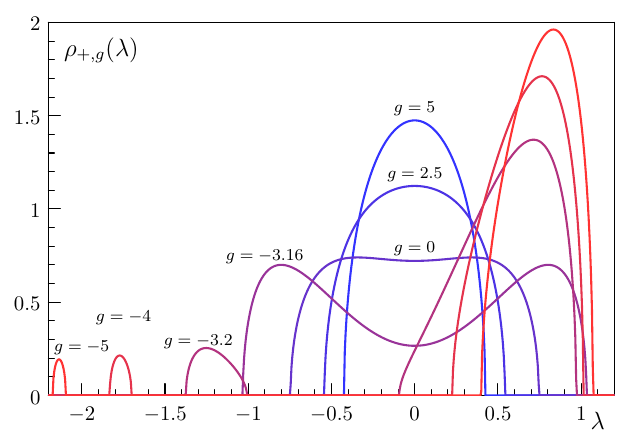}
\caption{\label{fig_10_density} Equilibrium densities in the \((1,0)\) geometries for various values of the coupling constant above and below the critical value.}
\end{figure}
Our work was originally motivated 
by the fact that Monte-Carlo simulations
showed a \emph{qualitatively} 
different behaviour. 
The simulations of Glaser and Barrett 
(and further simulations 
done by ourselves \cite{mauro_phd}) 
indicate that there is a 
(first order)
phase transition 
from a symmetric 1-cut solutions for 
\(g > g_{+,\mathrm{cr}}\)
to a broken symmetry 2-cut solution for 
\(g < g_{+,\mathrm{cr}}\).
One can simplify the simultaneous
nonlinear equations and  
reduce seven nonlinear equations for 
seven parameters, to four nonlinear equations
for four parameters.
From these we could not establish any relevant
new solutions by analytical means
and proceeded with numerical methods.
We did not search for broken symmetry 
1-cut solutions as there 
is no Monte-Carlo evidence.
In the 2-cut case there is Monte-Carlo 
evidence for broken symmetry solutions
and we used standard Newton-Raphson methods
to search for solutions. 
As starting points for the Newton-Raphson 
method we used approximate 
values for the boundaries of the 
support, that can be can be obtained 
from Monte-Carlo simulations.
In this way we found an additional broken symmetry solution with density
\(\rho^{\mathrm{2-cut, bs}}_{+,g}(\lambda)\) and positive first moment \(m_1>0\).
Note that \(\rho^{\mathrm{2-cut, bs}}_{+,g}(-\lambda)\) is a second solution
with \(m_1<0\). 
With this solution we confirm the Monte-Carlo
simulations qualitatively and quantitatively.\\ 
\begin{figure}
\includegraphics[width=0.49\textwidth]{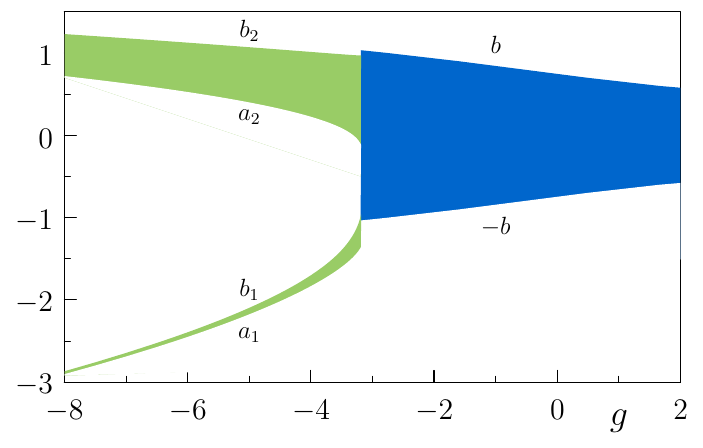}
\includegraphics[width=0.49\textwidth]{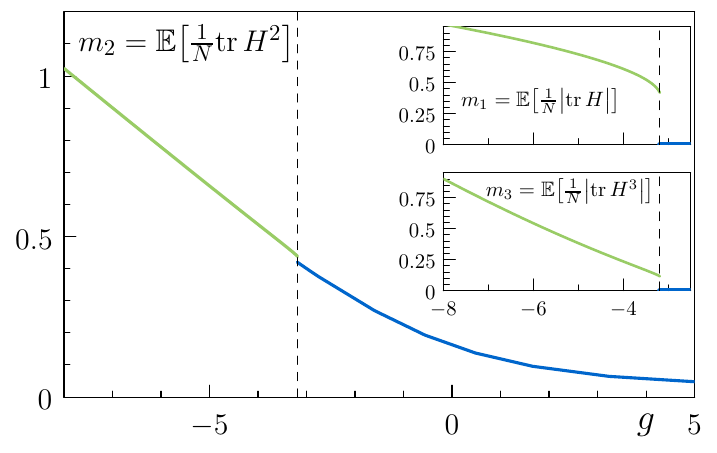}
\caption{\label{fig_support} Left: Support of the equilibrium density in the \((1,0)\) case where the support for the 
1-cut densities for \(g> g_{+,\mathrm{cr}}\) is shown in blue and the support of
the broken symmetry 2-cut densities for \(g> g_{+,\mathrm{cr}}\)
is shown in green.\\
Right: Dependence of the moments \(m_2\) (large graph), \(m_1\) (upper inset), and \(m_3\) (lower inset)
on the coupling constant. The critical value \(g_{+,\mathrm{cr}}\) is indicated by the vertical 
dashed lines. The jumps of all moments at the critical value indicate a 1st order phase transition.
}
\end{figure}
Let us establish 
that the broken symmetry 
solution is the relevant minimum of the free energy for suitable
coupling constants and establish any critical values where the
nature of the minimum changes.
In Figure~\ref{fig_free_energy} the free energies of the symmetric 1-cut, 
the symmetric 2-cut and the numerically found broken symmetry 2-cut solution
are plotted against the coupling constant \(g\).
As the equilibrium density is a minimum of the free energy one can read off
the relevant solutions. This reveals that 
the symmetric 2-cut solution is not relevant 
for any value of the
coupling constant, the symmetric 1-cut solution has minimal free energy
for \(g > g_{+,\mathrm{cr}}\), and the 
broken symmetry 2-cut solution for \(g < g_{+,\mathrm{cr}}\).
The critical value 
\( g_{+,\mathrm{cr}}\approx -3.187\) 
is established by the crossing of the two lines.\\
Figure~\ref{fig_10_density} shows the 
equilibrium densities for some values of the coupling constant above and below the
critical value.
%In
%Figure~\ref{fig_moments} shows the phase %transitions in the 
%moments
%\begin{figure}
%\includegraphics[width=0.9\textwidth]{moments_10.pdf}
%\caption{\label{fig_moments} Moments}
%\end{figure}
For completeness, the left graph in
Figure~\ref{fig_support} shows the boundaries 
of the
support in the \((1,0)\) case and the right
graph shows how the spectral moments depend on
the coupling constant. The jump at the critical
value \(g_{+,\mathrm{cr}}\) indicates a first-order phase transition.
 \\
Let us note that 
the mean density of states is given by
\begin{equation}
\mathbb{E}\left[\rho_{\Lambda}(\lambda) \right]=
\begin{cases}
    \rho_{+,g}^{\mathrm{1-cut}}(\lambda) &
    \text{for \(g>g_{+,\mathrm{cr}}\),}\\
    \frac{\rho_{+,g}^{\mathrm{2-cut, bs}}(\lambda)
        + \rho_{+,g}^{\mathrm{2-cut, bs}}(-\lambda)
    }{2} &
    \text{for \(g<g_{+,\mathrm{cr}}\).}
\end{cases}
\end{equation}
Furthermore our findings support that for \(g>g_{+,\mathrm{cr}}\)
the density of states converges to the equilibrium density
\( \rho_{\Lambda}(\lambda) \to \rho_{+,g}^{\mathrm{1-cut}}(\lambda)\)
in a weak sense as \(N\to \infty\) while this is not the case
for \(g<g_{+,\mathrm{cr}}\). In the latter case there are two accumulation points
and convergence can be recovered, e.g. by adding the constraint \(\mathrm{tr}\, H \ge 0\) 
to the ensemble.

\subsection{Monte-Carlo simulations}

\begin{figure}
\includegraphics[width=0.32\textwidth]{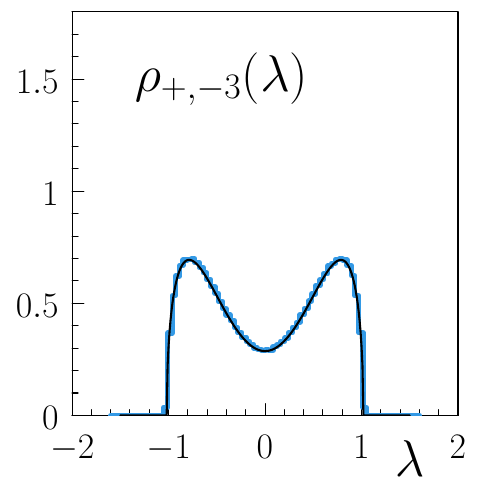}
\includegraphics[width=0.32\textwidth]{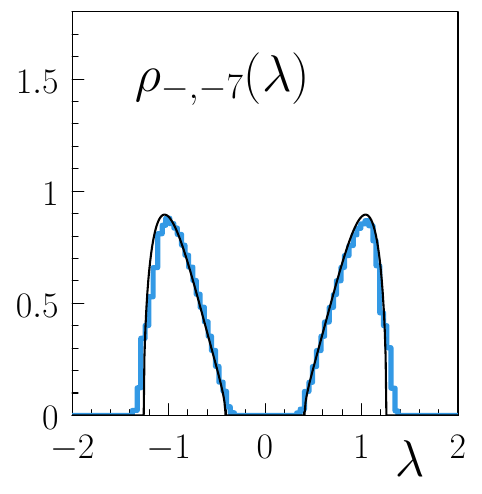}
\includegraphics[width=0.32\textwidth]{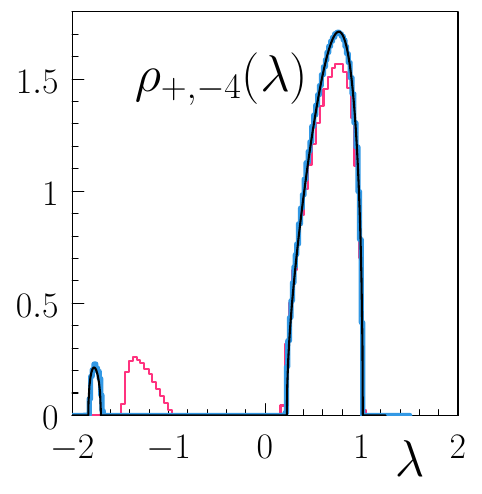}
\caption{\label{fig_MC} Comparison 
%%change ' Comparison of Monte-Carlo simulations ->
%% Comparison of configurations of the Monte-Carlo simulation
of configurations of the Monte-Carlo simulation for the equilibrium
density 
at matrix size \(N=1024\) 
(blue and red curves) with the 
theoretical equilibrium density 
(black) that we obtained using the 
Riemann-Hilbert approach. All plots are free 
of any free fitting parameters.\\
Left: 
1-cut spectral density in  the \((1,0)\) case  at coupling
constant \(g=-3\). The \((0,1)\) (not shown)
shows equally good agreement.
\\
Middle: Symmetric 2-cut spectral density 
in the \((0,1)\) case at coupling constant \(g=-7\).\\
Right:
2-cut spectral density with broken symmetry in the
\((1,0)\) case at coupling constant 
\(g=-4\). The blue and red curves correspond to
different starting configurations of the the 
Monte-Carlo simulation. For the blue case we 
started from a configuration close to the 
theoretical curve (black). For the red curve we 
started from evenly spaced eigenvalues and the 
Monte-Carlo simulation ran into a false configuration
with higher free energy than the blue curve.
}
\end{figure}

The theoretical equilibrium densities that we 
%%change have have -> have
have found above using the Riemann-Hilbert approach
are in very good agreement to the 
Monte-Carlo simulations by Barrett and Glaser. 
As the latter were done 
at relatively small matrix dimension we have redone
the relevant
Monte-Carlo simulations at 
increased matrix size \(N=1024\).
We implemented a Metropolis algorithm using the 
Random Fuzzy Library \cite{RFL}.
We refer to \cite{mauro_phd} for technical
details of the implementation.\\
Figure~\ref{fig_MC} compares the equilibrium 
densities 
%%change with with -> with 
with mean densities obtained through 
Monte-Carlo simulations. 
In the case of broken symmetries the 
Monte-Carlo simulation converge to one of the two
symmetry related equilibrium densities and we 
always show the one with \(m_1 \ge 0\).
The almost perfect  match between analytical 
results and 
Monte-Carlo simulation (with no fitting parameters)
is a strong 
validation of both the theoretical derivation
of equilibrium densities using the 
Riemann-Hilbert method and the Monte-Carlo 
method. It also shows that finite size 
effects are very small at 
the matrix size \(N=1024\) used in 
the simulation.
We should note one challenge for 
Monte-Carlo simulations of this type as we observe
in the broken symmetry case 
that the simulation can get stuck in a non-optimal
solution 
(see right graph in Figure~\ref{fig_MC}).
In this case we needed to start the simulation close 
to the theoretical curve in order to converge to the
optimal solution that minimizes the free energy.
Without sufficient knowledge of the true minimum
Monte-Carlo methods may in general have difficulties
with finding the true minimum. Indeed our 
simulations have never converged to 
%% change to the true minimum -> 
%% to what we now believe to
%% be the true minimum
%% 
what we now believe to be the true 
minimum before we had a theoretical prediction in 
our hands that allowed us to start sufficiently close. 
Improved Monte-Carlo algorithms such as Hamiltonian 
Monte-Carlo \cite{HMC} might be able to achieve the 
correct minimum. However, this possibility was not 
explored here.

\subsection{Implications to the Dirac spectrum}

In this paper we have focused on the spectral properties of the Hamiltonian
matrix \(H\) that defines the Dirac operators \eqref{Dirac}.
For completeness let us briefly state how to obtain the spectral densities for the
Dirac operators. For a given hermitian matrix \(H= U \Lambda U^\dagger\) with 
eigenvalues \(\lambda_n\) (\(n=1,\dots,N\)) and density of states 
\(\rho_{\Lambda}(\lambda)=
\frac{1}{N} \sum_{n=1}^N \delta(\lambda-\lambda_n)\) the eigenvalues of the corresponding
Dirac operators \(D_{\pm}\) are \(s_{\pm, mn}= \lambda_m \pm \lambda_n\) 
(\(m,n=1,\dots,N\)). The corresponding Dirac density of states is then
given by a convolution
\begin{equation}
    \rho_{D_\pm}(s) 
    = \frac{1}{N^2} \sum_{m,n=1}^N \delta(s- \lambda_m \mp \lambda_n)
    = 
    \int \rho_{\Lambda}(s \mp \lambda) \rho_{\Lambda}(\lambda) d\lambda \ .
\end{equation}
In the absence of symmetry breaking we have found evidence 
that \(\rho_\Lambda(\lambda)\) converges to the equilibrium density as \(N \to \infty\).
The same evidence can then be applied to the Dirac
density of states for the Dirac operators which then converges to the corresponding
convolution of the equilibrium densities \(\rho_{\pm, g}\). In the broken symmetry case
we have given evidence that there are two accumulation points -- and focussing on one
accumulation point (e.g. by requiring a positive trace) one again finds the same convolution.
We refer to the literature \cite{barrett2016, Glaser_2017, barrett2019} 
for a more detailed discussion to which we have nothing to add.

\section{Conclusions}

To conclude we have given a detailed account of the Riemann-Hilbert approach to
the spectral densities of two one-parameter families of ensembles
(referred to as \((1,0)\) and \((0,1)\)) 
of \(N\)-dimensional random matrices in the asymptotic limit
\(N \to \infty\). The latter describe crossovers or phase transitions in 
random fuzzy geometries in dependence on a coupling strength \(g\) as a 
toy model for quantum gravity.
We correct the previous pioneering approach by 
Khalkhali and Pagliaroli in two ways. First, we find some calculation errors that lead to 
almost perfect agreement with numerical Monte-Carlo simulation at finite dimension including 
the values of the critical coupling strengths.
Second, in the \((1,0)\) case we find a 1st-order phase transition 
accompanied by a symmetry breaking, such that for \(g> g_{+\mathrm{cr}}\approx -3.187\) 
one has a symmetric equilibrium density which is supported on one finite symmetric interval
while for \(g< g_{+\mathrm{cr}}\) 
one has an equilibrium density with broken symmetry that is supported on two finite
(non-symmetric) intervals. The moments of this density are not continuous at the critical
value (1st-order phase transition).
In the \((0,1)\) case we confirm that
there is a crossover (formally third-order phase transition)
from a symmetric equilibrium on one interval for \(g> g_{-,\mathrm{cr}}=-4\sqrt{2}\)
(this value differs from Khalkhali and Pagliaroli due to calculation errors on their side)
to a symmetric equilibrium on two intervals for \(g< g_{-,\mathrm{cr}}\).
In order to find the broken symmetry densities we need to generalize the 
Riemann-Hilbert approach of Khalkhali and Pagliaroli 
to allow for non-symmetric equilibrium densities.
We give the explicit analytical form of these densities which depends on
parameters that satisfy explicit non-linear equations.\\
In short we have given a complete theoretical picture of phase transitions
in these two ensembles. While some parameters that enter are defined by a set of non-linear 
implicit equations that can only be solved numerically, the derivation is completely analytical.
\\
We have not considered local fluctuations in the spectra in detail as a signature of quantum chaos \cite{QSOC}. Based on the overwhelming
evidence from more general random-matrix ensembles one expects that the spectral fluctuations 
in the bulk  (inside the support) have universal spectral fluctuations on the scale of the 
mean level spacing as predicted by the Gaussian Unitary Ensemble (GUE) \cite{mehta_book, Bonnet}. 
Mathematically rigorous approaches to universality
that use similar methods may be extended to the present context \cite{PS, deift2000}
\\
The Riemann-Hilbert method used here can in general not be applied to ensembles that depend on
more than one matrix. In more general \(p,q\) geometries one has \(2^{p+q-1}\) hermitian random 
matrices with a joint probability measure that gives rise to correlations. 
Still, 
the Riemann-Hilbert method and the results obtained here may be useful as a starting point
that builds in correlations (interactions between the various Coulomb gases for each matrix)
perturbatively. We leave this as an interesting future direction of research.

\acknowledgments
We would like to thank John Barrett and 
Liel 
%%change Lisa -> Liel
Glaser for many fruitful discussions 
throughout the project. SG thanks Gernot Akemann for
pointing to relevant random-matrix literature.

\appendix

\section{Explicit Borel transforms 
and 1-cut and 2-cut spectral measures 
for ensembles \(d\mu_{W}(H) = \frac{1}{Z_W} 
e^{- N \mathrm{tr}\ W(H)}\) 
and \(W(\lambda)=\sum_{k=1}^4 w_k \lambda^k\)}
\label{app_Wlambda}

Let
\(W(\lambda)= w_1 \lambda +w_2 \lambda^2+ w_3 \lambda^3
+ w_4 \lambda^4\) (with \(w_4>0\)) 
be a polynomial of order \(4=2M\). Any constant term \(w_0\) in the polynomial 
is irrelevant in this context (as it can be absorbed into the normalization of the
random-matrix ensemble), so we set \(w_0=0\) throughout.
In this appendix we give detailed account of the equilibrium measure
and the corresponding Borel transform for random-matrix ensembles defined by this potential.
The details are used in the main text of this work.
We assume (on physical grounds) that the number of cuts is 
either \(K=1\)
or \(K=2\) (the potential has at most two local minima).\\
We start with \(K=1\) where we write \(\Omega= [a,b]\)
and \(q=(z-a)(z-b)\). 
Introducing the symmetric combinations 
\begin{subequations}
    \begin{align}
        S_1=&a+b\\
        S_2=&ab
    \end{align}
\end{subequations}
and performing the
contour integration in \eqref{eq_Borel_Plemelj} one finds
\begin{subequations}
\begin{align}
    G_{\rho_W}(z)=&-\frac{w_1+2w_2z+3w_3z^2+4w_4 z^3
    }{2\pi i}
    \label{eq_Gz_1cut}
    \\%\left(w_1+2w_2z+3w_3z^2+4w_4 z^3\right)\nonumber\\
    &+\frac{\sqrt{q(z)}
    \left(2w_2+\frac{3S_1}{2}w_3 + \frac{3S_1^2 - 4  S_2}{2}w_4 +\left(3w_3 + 2w_4S_1\right)z+4w_4 z^2 \right)}{2 \pi i}\nonumber\\
     \rho_{W}(\lambda)= &
    \mathrm{Re} \lim_{\epsilon\to 0^+}
    G_{\rho_W}(\lambda+i\epsilon)\nonumber\\
    =&
    \begin{cases}
        \frac{\sqrt{q(-\lambda)}}{\pi}
        \left(w_2+ \frac{3S_1}{4}w_3 + \frac{3S_1^2-4S_2}{4}w_4 
        +\frac{3w_3+2w_4S_1}{2}\lambda+ 2w_4\lambda^2
        \right)
            & \text{for \(\lambda \in [a,b]\),}\\
        0 & \text{else.}
    \end{cases}
    %=& g_0 + g_1 z^{-1} +O\left(z^{-2}\right)
\end{align}
\end{subequations}
For the asymptotic expansion of \eqref{eq_Gz_1cut} let us first expand
\begin{align}
    \sqrt{q(z)}=\sqrt{z^2-S_1 z+ S_2}=& z\left(1 - \sum_{n=1}^\infty\frac{C_n}{z^n} \right)
\end{align}
\begin{subequations}
with the explicit leading coefficients
\begin{align}
    C_1=&\frac{S_1}{2} \\
    C_2=& \frac{S_1^2 -4S_2}{8} \\
    C_3=&\frac{S_1^3 -4 S_1 S_2}{16}\\
    C_4=&\frac{5S_1^4-24 S_1^2S_2+16S_2^2}{128}
\end{align}
\end{subequations}
This gives the asymptotic expansion
\begin{subequations}
\begin{align}
i \pi 
 G_{\rho_W}(z)
 =& \left(1-\sum_{n=1}^\infty \frac{C_n}{z^n}\right)
 \left(2 w_4z^3 +\frac{3w_3+4C_1 w_4}{2}z^2 +
 \frac{4(C_1^2+C_2)w_4+   3C_1 w_3 +2w_2}{2} z
 %2(C_1^2+C_2)w_4z + \frac{3}{2} C_1 w_3 z +w_2 z 
 \right)\nonumber\\
 &-  2w_4 z^3 -\frac{3}{2} w_3 z^2 -w_2z -\frac{w_1}{2}\\
 \sim& -
 \frac{4w_4 C_3+ (3w_3+4C_1 w_4) C_2+(4(C_1^2+C_2)w_4+ 
 3C_1 w_3 +2w_2)C_1 + w_1}{2}\nonumber\\
 & -\sum_{n=1}^\infty 
 \frac{
 2w_4 C_{n+3}+ \frac{3w_3+4C_1 w_4}{2} C_{n+2}+\frac{4(C_1^2+C_2)w_4+
 3C_1 w_3 +2w_2}{2}C_{n+1}}{z^n}\nonumber\\
 %\\
 %&- \frac{1}{z}\left( \right)
%=& g_0 + g_1 z^{-1} +O\left(z^{-2}\right)\\
%g_0=& \frac{i}{\pi}\left(\frac{w_1}{2} +\frac{s_1w_2}{2} + 
%\frac{(9s_1^2-12s_2)w_3}{16} 
%+\frac{(5s_1^3 -12s_1 s_2)w_4}{8}\right)\\
%g_1=&\frac{i}{\pi}\left(s_1^2-4s_2\right)\left(\frac{w_2}{8}+\frac{3s_1 w_3}
%{16}+ \frac{(15s_1^2-12s_2)w_4}{64}\right)&
\end{align}
\end{subequations}
For consistency with \(G_{\rho_W}(z) = \frac{i}{\pi z} + \mathcal{O}(z^{-2})\) one then has
the two conditions
\begin{subequations}
    \begin{align}
       2w_4 C_3+ \frac{3w_3+4C_1 w_4}{2} C_2+\frac{4(C_1^2+C_2)w_4+  
       3C_1 w_3 +2w_2}{2}C_1 + \frac{w_1}{2}= & 0,\\
       2w_4 C_{4}+ \frac{3w_3+4C_1 w_4}{2} C_{3}+\frac{4(C_1^2+C_2)w_4+  
       3C_1 w_3 +2w_2}{2}C_{2}=&1.
    \end{align}
\end{subequations}
These conditions
define \(a\) and \(b\) 
implicitly in terms of the coefficients \(w_k\). Note that 
existence and uniqueness of the equilibrium measure does not imply
that the conditions have a unique solution. It ensures that there is 
at least one solution among this and other multi-cut solutions 
and among the solutions there is a unique one that
has minimal free energy. \\
For the 2-cut case let
\(a_1<b_1<a_2<b_2\) such that the support is
\begin{equation}
    \Omega=[a_1,b_1] \cup [a_2,b_2]
\end{equation}
and
\begin{equation}
    q(z)=(z-a_1)(z-b_1)(z-a_2)(z-b_2).
\end{equation}
It is useful to introduce the symmetric combinations
\begin{subequations}
\begin{align}
    s_1=& a_1+b_1+a_2+b_2\\
    s_2=& a_1b_1+a_2b_2+ a_1a_2 + b_1 b_2 +a_1b_2 + b_1a_2\\
    s_3=& a_1b_1a_2 + a_1 b_1 b_2 + a_1 a_2 b_2 + b_1 a_2 b_2\\
    s_4=& a_1 b_1 a_2 b_2
\end{align}
\end{subequations}
such that \(q(z)=z^4-s_1z^3+ s_2z^2-s_3 z+s_4\).
Contour integration in \eqref{eq_Borel_Plemelj} one finds
\begin{subequations}
\begin{align}
    G_{\rho_W}(z)=&-\frac{w_1+2w_2z+3w_3z^2+4w_4 z^3
    }{2\pi i}
+\frac{\sqrt{q(z)}
    \left(  3 w_3 +2w_4 s_1 +4w_4 z\right)}{2 \pi i}    \label{eq_Gz_2cut}
    \\
     \rho_{W}(\lambda)
    =&
    \begin{cases}
        \frac{\sqrt{-q(\lambda)}}{\pi}
        \left(\frac{3}{2}w_3 + s_1 w_4 
        + 2 w_4 \lambda
        \right)
            & \text{for \(\lambda \in [a_2,b_2]\),}\\
        -\frac{\sqrt{-q(\lambda)}}{\pi}
        \left(\frac{3}{2}w_3 + s_1 w_4 
        + 2 w_4 \lambda
        \right)
            & \text{for \(\lambda \in [a_1,b_1]\),}\\
        0 & \text{else.}
    \end{cases}
\end{align}
\end{subequations}
The four interval boundaries of the support are defined by four consistency 
equations. Three of these follow from consistency requirements for 
\(G_{\rho_W}(z)\) as \(|z| \to \infty\). For this it is useful to 
first introduce the asymptotic expansion 
\begin{equation}
    \sqrt{q(z)} \sim z^2\left(1 -\sum_{n=1}^\infty c_n z^{-n} \right) \ .
\end{equation}
All coefficients \(c_n\) 
depend only on the boundary values of the support and this is best
expressed in terms of the symmetric combinations.
Explicitly, one obtains
%\begin{footnotesize}
\begin{subequations}
\begin{align}
    c_1=&\frac{s_1}{2},\\
    c_2=&\frac{s_1^2-4s_2}{8},\\
    c_3=& \frac{s_1^3-4s_1 s_2 +8 s_3}{16},\\
    c_4=& \frac{5s_1^4-24s_1^2 s_2+32s_1s_3+16 s_2^2-64 s_4}{128},\\
    c_5=& \frac{7 s_1^5 -40 s_1^3s_2+48 s_1^2 s3+
    48  s_1s_2^2-64s_1s_4-64s_2s_3}{256},\\
    c_6 = & \frac{21 s_1^6-140 s_1^4 s_2+160s_1^3s_3+240 s_1^2s_2^2
    -192s_1^2 s4-384 s_1 s_2 s_3-64 s_2^3 
    }{1024}\nonumber\\
    &\quad +\frac{256 s_2 s_4+128s_3^2
    }{1024},\\
    c_7 = & \frac{33 s_1^7-252s_1^5 s_2+280 s_1^4 s_3+560 s_1^3s_2^2
    -320 s_1^3s_4-960 s_1^2s_2 s_3-320 s_1 s_2^3
    }{2048}\nonumber\\
    &\quad +\frac{768 s_1s_2 s_4+384s_1s_3^2+384 s_2^2 s_3 -512 s_3s_4}{2048},
\end{align}
\end{subequations}
%\end{footnotesize}
where we included some coefficients that are needed in Appendix~\ref{app_2cut_conditions}.
Now we can write
\begin{align}
     i \pi G(z)=&\left(1-\sum_{n=1}^\infty 
    \frac{c_n}{z^n}\right)\left(2 w_4 z^3 +
    \frac{4 c_1 w_4 + 3w_3 }{2} z^2
    \right)\nonumber 
    \\
    &\qquad -2w_4z^3 -\frac{3}{2} w_3z^2  -w_2z -\frac{w_1}{2}\nonumber 
    \\
    \sim &\quad
    -z^1\left(2c_2 w_4 
    + c_1\frac{4 c_1 w_4 + 3w_3 }{2} 
    +w_2\right)\nonumber \\
    & \quad -
    z^0\left(2c_3 w_4
    + c_2\frac{4 c_1 w_4 + 3w_3 }{2} 
    +\frac{w_1}{2}\right)
   \nonumber \\
    & \quad -
    \sum_{n=1}^\infty
    z^{-n}\left(    
    2c_{n+3} w_4
    + c_{n+2}\frac{4 c_1 w_4 + 3w_3 }{2} 
    \right)\ . 
    \label{eq_Gz_2cu_as}
\end{align}
Consistency of the asymptotic expansion requires the three conditions 
\begin{subequations}
    \begin{align}
    2c_2 w_4 
    + c_1\frac{4 c_1 w_4 + 3w_3 }{2} 
    +w_2=&0\\
    2c_3 w_4
    + c_2\frac{4 c_1 w_4 + 3w_3 }{2} =&0\\
    2c_{4} w_4
    + c_{3}\frac{4 c_1 w_4 + 3w_3 }{2} =&1\ .
    \end{align}
    One further condition follows from the variational approach.
    The Lagrange multiplier on the right-hand side 
    of \eqref{eq_cond_U} neads to be the same constant in 
    both intervals \cite{deift2000} which results in
    \begin{equation}
     \int_{b_1}^{a_2} \sqrt{q(x)}
        \left(
           4 w_4x++ 4 w_4 c_1+3w_3
        \right) dx=0\ .
    \end{equation}
    \label{2cut_conditions_gen_W}
\end{subequations}
These four non-linear conditions implicitly define 
the boundaries \(a_1<b_1<a_2<b_2\)
of the support. 

\section{Explicit conditions on the support in the two interval case: 
\((1,0)\) case}
\label{app_2cut_conditions}

In the \((1,0)\) geometries the coefficients, substituting the coefficients 
\(w_1=8m_3+4g m_1\), \(w_2=12m_2+2g\), \(w_3=8m_1\)
and \(w_4=2\) of the effective 
potential in \eqref{2cut_conditions_gen_W}
gives
\begin{subequations}
    \begin{align}
        4c_2+4c_1(c_1+3m_1) + 2(g_2+6m_2)=& 0
        \label{cond1}
        \\
        4c_3+4c_2(c_1+3m_1) + 2(g_2m_1+2m_3)=&0
        \label{cond2}\\
        4c_{4}+4c_{3}(c_1+3m_1)=&1
        \label{cond3}\\      
        \int_{b_1}^{a_2} \sqrt{q(x)}
        \left(
           x+c_1+3m_1
        \right) dx=&0
        \label{cond7}
        \end{align}
        Here the parameters \(m_k\) (\(k=1,2,3\))
        are the first three spectral moments.
        The spectral moments can be read of directly 
        from the correspondig order in the asymptotic expansion \eqref{eq_Gz_2cu_as}.
        Consistency then leads to the additional three conditions
        \begin{align}
        4c_{5}+4c_{4}(c_1+3m_1)=&m_1
        \label{cond4}\\
        4c_{6}+4c_{5}(c_1+3m_1)=&m_2
        \label{cond5}\\
        4c_{7}+4c_{6}(c_1+3m_1)=&m_3\ .
        \label{cond6}
    \end{align}
    \label{allconditions}
\end{subequations}
Altogether these are seven equations for seven parameters.
The last three equations can be used to express the
spectral moments in term of the boundaries of the support. 
This leaves effectively four nonlinear equations for four 
parameters. One should note that there may be more 
than one solution to these equations and one needs to 
compare the free energies of the corresponding densities
to find the equilibrium density.


\begin{thebibliography}{29}%
\makeatletter
\providecommand \@ifxundefined [1]{%
 \@ifx{#1\undefined}
}%
\providecommand \@ifnum [1]{%
 \ifnum #1\expandafter \@firstoftwo
 \else \expandafter \@secondoftwo
 \fi
}%
\providecommand \@ifx [1]{%
 \ifx #1\expandafter \@firstoftwo
 \else \expandafter \@secondoftwo
 \fi
}%
\providecommand \natexlab [1]{#1}%
\providecommand \enquote  [1]{``#1''}%
\providecommand \bibnamefont  [1]{#1}%
\providecommand \bibfnamefont [1]{#1}%
\providecommand \citenamefont [1]{#1}%
\providecommand \href@noop [0]{\@secondoftwo}%
\providecommand \href [0]{\begingroup \@sanitize@url \@href}%
\providecommand \@href[1]{\@@startlink{#1}\@@href}%
\providecommand \@@href[1]{\endgroup#1\@@endlink}%
\providecommand \@sanitize@url [0]{\catcode `\\12\catcode `\$12\catcode
  `\&12\catcode `\#12\catcode `\^12\catcode `\_12\catcode `\%12\relax}%
\providecommand \@@startlink[1]{}%
\providecommand \@@endlink[0]{}%
\providecommand \url  [0]{\begingroup\@sanitize@url \@url }%
\providecommand \@url [1]{\endgroup\@href {#1}{\urlprefix }}%
\providecommand \urlprefix  [0]{URL }%
\providecommand \Eprint [0]{\href }%
\providecommand \doibase [0]{http://dx.doi.org/}%
\providecommand \selectlanguage [0]{\@gobble}%
\providecommand \bibinfo  [0]{\@secondoftwo}%
\providecommand \bibfield  [0]{\@secondoftwo}%
\providecommand \translation [1]{[#1]}%
\providecommand \BibitemOpen [0]{}%
\providecommand \bibitemStop [0]{}%
\providecommand \bibitemNoStop [0]{.\EOS\space}%
\providecommand \EOS [0]{\spacefactor3000\relax}%
\providecommand \BibitemShut  [1]{\csname bibitem#1\endcsname}%
\let\auto@bib@innerbib\@empty
%</preamble>
\bibitem [{\citenamefont {Barrett}\ and\ \citenamefont
  {Glaser}(2016)}]{barrett2016}%
  \BibitemOpen
  \bibfield  {author} {\bibinfo {author} {\bibfnamefont {J.~W.}\ \bibnamefont
  {Barrett}}\ and\ \bibinfo {author} {\bibfnamefont {L.}~\bibnamefont
  {Glaser}},\ }\href@noop {} {\bibfield  {journal} {\bibinfo  {journal}
  {J.~Phys.~A}\ }\textbf {\bibinfo {volume} {49}},\ \bibinfo {pages} {245001}
  (\bibinfo {year} {2016})}\BibitemShut {NoStop}%
\bibitem [{\citenamefont {Barrett}(2015)}]{barrett2015}%
  \BibitemOpen
  \bibfield  {author} {\bibinfo {author} {\bibfnamefont {J.~W.}\ \bibnamefont
  {Barrett}},\ }\href {\doibase 10.1063/1.4927224} {\bibfield  {journal}
  {\bibinfo  {journal} {J.~Math.~Phys.}\ }\textbf {\bibinfo {volume} {56}},\
  \bibinfo {pages} {082301} (\bibinfo {year} {2015})}\BibitemShut {NoStop}%
\bibitem [{\citenamefont {Mehta}(2004)}]{mehta_book}%
  \BibitemOpen
  \bibfield  {author} {\bibinfo {author} {\bibfnamefont {M.~L.}\ \bibnamefont
  {Mehta}},\ }\href@noop {} {\emph {\bibinfo {title} {Random Matrices}}}\
  (\bibinfo  {publisher} {Academic Press},\ \bibinfo {year} {2004})\BibitemShut
  {NoStop}%
\bibitem [{\citenamefont {Forrester}(2010)}]{forrester}%
  \BibitemOpen
  \bibfield  {author} {\bibinfo {author} {\bibfnamefont {P.~J.}\ \bibnamefont
  {Forrester}},\ }\href@noop {} {\emph {\bibinfo {title} {Log-Gases and Random
  Matrices}}}\ (\bibinfo  {publisher} {Princeton University Press},\ \bibinfo
  {year} {2010})\BibitemShut {NoStop}%
\bibitem [{\citenamefont {Akemann}\ \emph {et~al.}(2011)\citenamefont
  {Akemann}, \citenamefont {Baik},\ and\ \citenamefont
  {Di~Francesco}}]{handbook}%
  \BibitemOpen
  \bibinfo {editor} {\bibfnamefont {G.}~\bibnamefont {Akemann}}, \bibinfo
  {editor} {\bibfnamefont {J.}~\bibnamefont {Baik}}, \ and\ \bibinfo {editor}
  {\bibfnamefont {P.}~\bibnamefont {Di~Francesco}},\ eds.,\ \href@noop {}
  {\emph {\bibinfo {title} {The Oxford Handbook of Random Matrix Theory}}}\
  (\bibinfo  {publisher} {Oxford University Press},\ \bibinfo {year}
  {2011})\BibitemShut {NoStop}%
\bibitem [{\citenamefont {Haake}\ \emph {et~al.}(2018)\citenamefont {Haake},
  \citenamefont {Gnutzmann},\ and\ \citenamefont {Ku{\'s}}}]{QSOC}%
  \BibitemOpen
  \bibfield  {author} {\bibinfo {author} {\bibfnamefont {F.}~\bibnamefont
  {Haake}}, \bibinfo {author} {\bibfnamefont {S.}~\bibnamefont {Gnutzmann}}, \
  and\ \bibinfo {author} {\bibfnamefont {M.}~\bibnamefont {Ku{\'s}}},\
  }\href@noop {} {\emph {\bibinfo {title} {Quantum Signatures of Chaos}}},\
  \bibinfo {edition} {4th}\ ed.\ (\bibinfo  {publisher} {Springer},\ \bibinfo
  {year} {2018})\BibitemShut {NoStop}%
\bibitem [{\citenamefont {Glaser}(2017)}]{Glaser_2017}%
  \BibitemOpen
  \bibfield  {author} {\bibinfo {author} {\bibfnamefont {L.}~\bibnamefont
  {Glaser}},\ }\href@noop {} {\bibfield  {journal} {\bibinfo  {journal}
  {J.~Phys.~A}\ }\textbf {\bibinfo {volume} {50}},\ \bibinfo {pages} {275201}
  (\bibinfo {year} {2017})}\BibitemShut {NoStop}%
\bibitem [{\citenamefont {Barrett}\ \emph {et~al.}(2019)\citenamefont
  {Barrett}, \citenamefont {Druce},\ and\ \citenamefont
  {Glaser}}]{barrett2019}%
  \BibitemOpen
  \bibfield  {author} {\bibinfo {author} {\bibfnamefont {J.~W.}\ \bibnamefont
  {Barrett}}, \bibinfo {author} {\bibfnamefont {P.}~\bibnamefont {Druce}}, \
  and\ \bibinfo {author} {\bibfnamefont {L.}~\bibnamefont {Glaser}},\
  }\href@noop {} {\bibfield  {journal} {\bibinfo  {journal} {J.~Phys.~A}\
  }\textbf {\bibinfo {volume} {52}},\ \bibinfo {pages} {275203} (\bibinfo
  {year} {2019})}\BibitemShut {NoStop}%
\bibitem [{\citenamefont {Khalkhali}\ and\ \citenamefont
  {Pagliaroli}(2020)}]{khalkhali2020}%
  \BibitemOpen
  \bibfield  {author} {\bibinfo {author} {\bibfnamefont {M.}~\bibnamefont
  {Khalkhali}}\ and\ \bibinfo {author} {\bibfnamefont {N.}~\bibnamefont
  {Pagliaroli}},\ }\href@noop {} {\bibfield  {journal} {\bibinfo  {journal}
  {J.~Phys.~A}\ }\textbf {\bibinfo {volume} {54}},\ \bibinfo {pages} {035202}
  (\bibinfo {year} {2020})}\BibitemShut {NoStop}%
\bibitem [{\citenamefont {Khalkhali}\ and\ \citenamefont
  {Pagliaroli}(2022)}]{khalkhali2022}%
  \BibitemOpen
  \bibfield  {author} {\bibinfo {author} {\bibfnamefont {M.}~\bibnamefont
  {Khalkhali}}\ and\ \bibinfo {author} {\bibfnamefont {N.}~\bibnamefont
  {Pagliaroli}},\ }\href@noop {} {\bibfield  {journal} {\bibinfo  {journal}
  {J.~Math.~Phys.}\ }\textbf {\bibinfo {volume} {63}},\ \bibinfo {pages}
  {053504} (\bibinfo {year} {2022})}\BibitemShut {NoStop}%
\bibitem [{\citenamefont {Hessam}\ \emph
  {et~al.}(2022{\natexlab{a}})\citenamefont {Hessam}, \citenamefont
  {Khalkhali},\ and\ \citenamefont {Pagliaroli}}]{hessam2022-bootstrapping}%
  \BibitemOpen
  \bibfield  {author} {\bibinfo {author} {\bibfnamefont {H.}~\bibnamefont
  {Hessam}}, \bibinfo {author} {\bibfnamefont {M.}~\bibnamefont {Khalkhali}}, \
  and\ \bibinfo {author} {\bibfnamefont {N.}~\bibnamefont {Pagliaroli}},\
  }\href@noop {} {\bibfield  {journal} {\bibinfo  {journal} {J.~Phys.~A}\
  }\textbf {\bibinfo {volume} {55}},\ \bibinfo {pages} {335204} (\bibinfo
  {year} {2022}{\natexlab{a}})}\BibitemShut {NoStop}%
\bibitem [{\citenamefont {Hessam}\ \emph
  {et~al.}(2022{\natexlab{b}})\citenamefont {Hessam}, \citenamefont
  {Khalkhali}, \citenamefont {Pagliaroli},\ and\ \citenamefont
  {Verhoeven}}]{hessam2022-review}%
  \BibitemOpen
  \bibfield  {author} {\bibinfo {author} {\bibfnamefont {H.}~\bibnamefont
  {Hessam}}, \bibinfo {author} {\bibfnamefont {M.}~\bibnamefont {Khalkhali}},
  \bibinfo {author} {\bibfnamefont {N.}~\bibnamefont {Pagliaroli}}, \ and\
  \bibinfo {author} {\bibfnamefont {L.~S.}\ \bibnamefont {Verhoeven}},\
  }\href@noop {} {\bibfield  {journal} {\bibinfo  {journal} {J.~Phys.~A}\
  }\textbf {\bibinfo {volume} {55}},\ \bibinfo {pages} {413002} (\bibinfo
  {year} {2022}{\natexlab{b}})}\BibitemShut {NoStop}%
\bibitem [{\citenamefont {Azarfar}\ and\ \citenamefont
  {Khalkhali}(2024)}]{azarfar2024random}%
  \BibitemOpen
  \bibfield  {author} {\bibinfo {author} {\bibfnamefont {S.}~\bibnamefont
  {Azarfar}}\ and\ \bibinfo {author} {\bibfnamefont {M.}~\bibnamefont
  {Khalkhali}},\ }\href@noop {} {\bibfield  {journal} {\bibinfo  {journal}
  {Annales de l’Institut H.~Poincar{\'e} D}\ }\textbf {\bibinfo {volume}
  {11}},\ \bibinfo {pages} {409} (\bibinfo {year} {2024})}\BibitemShut
  {NoStop}%
\bibitem [{\citenamefont {Barrett}(2024)}]{barrett2024fermion}%
  \BibitemOpen
  \bibfield  {author} {\bibinfo {author} {\bibfnamefont {J.~W.}\ \bibnamefont
  {Barrett}},\ }\href@noop {} {\bibfield  {journal} {\bibinfo  {journal}
  {J.~Phys.~A}\ }\textbf {\bibinfo {volume} {57}},\ \bibinfo {pages} {455201}
  (\bibinfo {year} {2024})}\BibitemShut {NoStop}%
\bibitem [{\citenamefont {Khalkhali}\ \emph {et~al.}(2025)\citenamefont
  {Khalkhali}, \citenamefont {Pagliaroli},\ and\ \citenamefont
  {Verhoeven}}]{khalkhali2025large}%
  \BibitemOpen
  \bibfield  {author} {\bibinfo {author} {\bibfnamefont {M.}~\bibnamefont
  {Khalkhali}}, \bibinfo {author} {\bibfnamefont {N.}~\bibnamefont
  {Pagliaroli}}, \ and\ \bibinfo {author} {\bibfnamefont {L.~S.}\ \bibnamefont
  {Verhoeven}},\ }\href@noop {} {\bibfield  {journal} {\bibinfo  {journal}
  {J.~Math.~Phys.}\ }\textbf {\bibinfo {volume} {66}},\ \bibinfo {pages}
  {053502} (\bibinfo {year} {2025})}\BibitemShut {NoStop}%
\bibitem [{\citenamefont {Jurkiewicz}(1991)}]{jurkiewicz}%
  \BibitemOpen
  \bibfield  {author} {\bibinfo {author} {\bibfnamefont {J.}~\bibnamefont
  {Jurkiewicz}},\ }\href@noop {} {\bibfield  {journal} {\bibinfo  {journal}
  {Phys.~Lett.~B}\ }\textbf {\bibinfo {volume} {261}},\ \bibinfo {pages} {260}
  (\bibinfo {year} {1991})}\BibitemShut {NoStop}%
\bibitem [{\citenamefont {Ambj{\o}rn}\ \emph {et~al.}(1993)\citenamefont
  {Ambj{\o}rn}, \citenamefont {Chekhov}, \citenamefont {Kristjansen},\ and\
  \citenamefont {Makeenko}}]{ambjorn1993matrix}%
  \BibitemOpen
  \bibfield  {author} {\bibinfo {author} {\bibfnamefont {J.}~\bibnamefont
  {Ambj{\o}rn}}, \bibinfo {author} {\bibfnamefont {L.}~\bibnamefont {Chekhov}},
  \bibinfo {author} {\bibfnamefont {C.}~\bibnamefont {Kristjansen}}, \ and\
  \bibinfo {author} {\bibfnamefont {Y.}~\bibnamefont {Makeenko}},\ }\href@noop
  {} {\bibfield  {journal} {\bibinfo  {journal} {Nucl.~Phys.~B}\ }\textbf
  {\bibinfo {volume} {404}},\ \bibinfo {pages} {127} (\bibinfo {year}
  {1993})}\BibitemShut {NoStop}%
\bibitem [{\citenamefont {Akemann}\ and\ \citenamefont
  {Ambj{\o}rn}(1996)}]{akemann1996JPA}%
  \BibitemOpen
  \bibfield  {author} {\bibinfo {author} {\bibfnamefont {G.}~\bibnamefont
  {Akemann}}\ and\ \bibinfo {author} {\bibfnamefont {J.}~\bibnamefont
  {Ambj{\o}rn}},\ }\href@noop {} {\bibfield  {journal} {\bibinfo  {journal}
  {J.~Phys.~A}\ }\textbf {\bibinfo {volume} {29}},\ \bibinfo {pages} {L555}
  (\bibinfo {year} {1996})}\BibitemShut {NoStop}%
\bibitem [{\citenamefont {Akemann}(1996)}]{akemann1996NPB}%
  \BibitemOpen
  \bibfield  {author} {\bibinfo {author} {\bibfnamefont {G.}~\bibnamefont
  {Akemann}},\ }\href@noop {} {\bibfield  {journal} {\bibinfo  {journal}
  {Nucl.~Phys.~B}\ }\textbf {\bibinfo {volume} {482}},\ \bibinfo {pages} {403}
  (\bibinfo {year} {1996})}\BibitemShut {NoStop}%
\bibitem [{\citenamefont {Bonnet}\ \emph {et~al.}(2000)\citenamefont {Bonnet},
  \citenamefont {David},\ and\ \citenamefont {Eynard}}]{Bonnet}%
  \BibitemOpen
  \bibfield  {author} {\bibinfo {author} {\bibfnamefont {G.}~\bibnamefont
  {Bonnet}}, \bibinfo {author} {\bibfnamefont {F.}~\bibnamefont {David}}, \
  and\ \bibinfo {author} {\bibfnamefont {B.}~\bibnamefont {Eynard}},\
  }\href@noop {} {\bibfield  {journal} {\bibinfo  {journal} {J.~Phys.~A}\
  }\textbf {\bibinfo {volume} {33}},\ \bibinfo {pages} {6739} (\bibinfo {year}
  {2000})}\BibitemShut {NoStop}%
\bibitem [{\citenamefont {Deift}(2000)}]{deift2000}%
  \BibitemOpen
  \bibfield  {author} {\bibinfo {author} {\bibfnamefont {P.}~\bibnamefont
  {Deift}},\ }\href@noop {} {\emph {\bibinfo {title} {Orthogonal Polynomials
  and Random Matrices: A Riemann-Hilbert Approach}}},\ Vol.~\bibinfo {volume}
  {3}\ (\bibinfo  {publisher} {American Mathematical Soc.},\ \bibinfo {year}
  {2000})\BibitemShut {NoStop}%
\bibitem [{\citenamefont {Iso}\ and\ \citenamefont {Kavalov}(1997)}]{iso1997}%
  \BibitemOpen
  \bibfield  {author} {\bibinfo {author} {\bibfnamefont {S.}~\bibnamefont
  {Iso}}\ and\ \bibinfo {author} {\bibfnamefont {A.}~\bibnamefont {Kavalov}},\
  }\href@noop {} {\bibfield  {journal} {\bibinfo  {journal} {Nucl.~Phys.~B}\
  }\textbf {\bibinfo {volume} {501}},\ \bibinfo {pages} {670} (\bibinfo {year}
  {1997})}\BibitemShut {NoStop}%
\bibitem [{\citenamefont {Boutet~de Monvel}\ \emph {et~al.}(1995)\citenamefont
  {Boutet~de Monvel}, \citenamefont {Pastur},\ and\ \citenamefont
  {Shcherbina}}]{BPS}%
  \BibitemOpen
  \bibfield  {author} {\bibinfo {author} {\bibfnamefont {A.}~\bibnamefont
  {Boutet~de Monvel}}, \bibinfo {author} {\bibfnamefont {L.}~\bibnamefont
  {Pastur}}, \ and\ \bibinfo {author} {\bibfnamefont {M.}~\bibnamefont
  {Shcherbina}},\ }\href@noop {} {\bibfield  {journal} {\bibinfo  {journal}
  {J.~Stat.~Phys.}\ }\textbf {\bibinfo {volume} {79}},\ \bibinfo {pages} {585}
  (\bibinfo {year} {1995})}\BibitemShut {NoStop}%
\bibitem [{\citenamefont {Johansson}(1996)}]{Johansson}%
  \BibitemOpen
  \bibfield  {author} {\bibinfo {author} {\bibfnamefont {K.}~\bibnamefont
  {Johansson}},\ }\href@noop {} {\bibfield  {journal} {\bibinfo  {journal}
  {Duke Math.~J.}\ }\textbf {\bibinfo {volume} {91}},\ \bibinfo {pages} {151}
  (\bibinfo {year} {1996})}\BibitemShut {NoStop}%
\bibitem [{\citenamefont {Wigner}(1955)}]{wigner1955}%
  \BibitemOpen
  \bibfield  {author} {\bibinfo {author} {\bibfnamefont {E.}~\bibnamefont
  {Wigner}},\ }\href@noop {} {\bibfield  {journal} {\bibinfo  {journal}
  {Ann.~Math.}\ }\textbf {\bibinfo {volume} {62}},\ \bibinfo {pages} {548}
  (\bibinfo {year} {1955})}\BibitemShut {NoStop}%
\bibitem [{\citenamefont {D'Arcangelo}(2022)}]{mauro_phd}%
  \BibitemOpen
  \bibfield  {author} {\bibinfo {author} {\bibfnamefont {M.}~\bibnamefont
  {D'Arcangelo}},\ }\emph {\bibinfo {title} {Numerical simulation of random
  Dirac operators}},\ \href@noop {} {\bibinfo {type} {Phd thesis}},\ \bibinfo
  {school} {University of Nottingham} (\bibinfo {year} {2022})\BibitemShut
  {NoStop}%
\bibitem [{\citenamefont {D'Arcangelo}(2020)}]{RFL}%
  \BibitemOpen
  \bibfield  {author} {\bibinfo {author} {\bibfnamefont {M.}~\bibnamefont
  {D'Arcangelo}},\ }\href@noop {} {\emph {\bibinfo {title} {Random Fuzzy
  Library}}} (\bibinfo {year} {2020}),\ \bibinfo {note} {available at
  \url{https://github.com/darcangelomauro/RFL}}\BibitemShut {NoStop}%
\bibitem [{\citenamefont {Duane}\ \emph {et~al.}(1987)\citenamefont {Duane},
  \citenamefont {Kennedy}, \citenamefont {Pendleton},\ and\ \citenamefont
  {Roweth}}]{HMC}%
  \BibitemOpen
  \bibfield  {author} {\bibinfo {author} {\bibfnamefont {S.}~\bibnamefont
  {Duane}}, \bibinfo {author} {\bibfnamefont {A.}~\bibnamefont {Kennedy}},
  \bibinfo {author} {\bibfnamefont {B.~J.}\ \bibnamefont {Pendleton}}, \ and\
  \bibinfo {author} {\bibfnamefont {D.}~\bibnamefont {Roweth}},\ }\href
  {\doibase https://doi.org/10.1016/0370-2693(87)91197-X} {\bibfield  {journal}
  {\bibinfo  {journal} {Physics Letters B}\ }\textbf {\bibinfo {volume}
  {195}},\ \bibinfo {pages} {216} (\bibinfo {year} {1987})}\BibitemShut
  {NoStop}%
\bibitem [{\citenamefont {Pastur}\ and\ \citenamefont {Shcherbina}(1997)}]{PS}%
  \BibitemOpen
  \bibfield  {author} {\bibinfo {author} {\bibfnamefont {L.}~\bibnamefont
  {Pastur}}\ and\ \bibinfo {author} {\bibfnamefont {M.}~\bibnamefont
  {Shcherbina}},\ }\href@noop {} {\bibfield  {journal} {\bibinfo  {journal}
  {J.~Stat.~Phys.}\ }\textbf {\bibinfo {volume} {86}},\ \bibinfo {pages} {109}
  (\bibinfo {year} {1997})}\BibitemShut {NoStop}%
\end{thebibliography}
\end{document}